%% file: 1AGabEW.tex
\documentclass[12pt]{amsart}

\def\aw{{\widetilde\alpha}}

\def\5d{{\dot\varepsilon}}

\def\m{\mu}

\def\n{\nu}

\def\1{{\dot 1}}
\def\2{{\dot 2}}

\def\8{\infty}

\def\0{\emptyset}

\def\pa{\partial}

\def\({\left(}
\def\){\right)}
\def\[{\left[}
\def\]{\right]}

\def\={x^\n{}\!,_\aw \[\pa_\m,\pa_\n\]x^\aw=0}

\def\sl{\textsl}
\def\bf{\textbf}

\def\eeq{\end{equation}}

\def\beqA{\begin{equation}\label{E:A}}
\def\beqB{\begin{equation}\label{E:B}}
\def\beqC{\begin{equation}\label{E:C}}
\def\beqD{\begin{equation}\label{E:D}}
\def\beqE{\begin{equation}\label{E:E}}
\def\beqF{\begin{equation}\label{E:F}}
\def\beqG{\begin{equation}\label{E:G}}
\def\beqH{\begin{equation}\label{E:H}}
\def\beqI{\begin{equation}\label{E:I}}
\def\beqJ{\begin{equation}\label{E:J}}
\def\beqK{\begin{equation}\label{E:K}}
\def\beqL{\begin{equation}\label{E:L}}
\def\beqM{\begin{equation}\label{E:M}}
\def\beqN{\begin{equation}\label{E:N}}
\def\beqO{\begin{equation}\label{E:O}}
\def\beqP{\begin{equation}\label{E:P}}
\def\beqQ{\begin{equation}\label{E:Q}}
\def\beqR{\begin{equation}\label{E:R}}
\def\beqS{\begin{equation}\label{E:S}}
\def\beqT{\begin{equation}\label{E:T}}
\def\beqU{\begin{equation}\label{E:U}}
\def\beqV{\begin{equation}\label{E:V}}
\def\beqW{\begin{equation}\label{E:W}}
\def\beqX{\begin{equation}\label{E:X}}
\def\beqY{\begin{equation}\label{E:Y}}
\def\beqZ{\begin{equation}\label{E:Z}}

\def\beqa{\begin{equation}\label{E:a}}
\def\beqb{\begin{equation}\label{E:b}}
\def\beqc{\begin{equation}\label{E:c}}
\def\beqd{\begin{equation}\label{E:d}}
\def\beqe{\begin{equation}\label{E:e}}
\def\beqf{\begin{equation}\label{E:f}}
\def\beqg{\begin{equation}\label{E:g}}
\def\beqh{\begin{equation}\label{E:h}}
\def\beqi{\begin{equation}\label{E:i}}
\def\beqj{\begin{equation}\label{E:j}}
\def\beqk{\begin{equation}\label{E:k}}
\def\beql{\begin{equation}\label{E:l}}
\def\beqm{\begin{equation}\label{E:m}}
\def\beqn{\begin{equation}\label{E:n}}
\def\beqo{\begin{equation}\label{E:o}}
\def\beqp{\begin{equation}\label{E:p}}
\def\beqq{\begin{equation}\label{E:q}}
\def\beqr{\begin{equation}\label{E:r}}
\def\beqs{\begin{equation}\label{E:s}}
\def\beqt{\begin{equation}\label{E:t}}
\def\bequ{\begin{equation}\label{E:u}}
\def\beqv{\begin{equation}\label{E:v}}
\def\beqw{\begin{equation}\label{E:w}}
\def\beqx{\begin{equation}\label{E:x}}
\def\beqy{\begin{equation}\label{E:y}}
\def\beqz{\begin{equation}\label{E:z}}

\def\beqAA{\begin{equation}\label{E:AA}}
\def\beqBB{\begin{equation}\label{E:BB}}
\def\beqCC{\begin{equation}\label{E:CC}}
\def\beqDD{\begin{equation}\label{E:DD}}
\def\beqEE{\begin{equation}\label{E:EE}}
\def\beqFF{\begin{equation}\label{E:FF}}
\def\beqGG{\begin{equation}\label{E:GG}}
\def\beqHH{\begin{equation}\label{E:HH}}
\def\beqII{\begin{equation}\label{E:II}}
\def\beqJJ{\begin{equation}\label{E:JJ}}
\def\beqKK{\begin{equation}\label{E:KK}}
\def\beqLL{\begin{equation}\label{E:LL}}
\def\beqMM{\begin{equation}\label{E:MM}}
\def\beqNN{\begin{equation}\label{E:NN}}
\def\beqOO{\begin{equation}\label{E:OO}}
\def\beqPP{\begin{equation}\label{E:PP}}
\def\beqQQ{\begin{equation}\label{E:QQ}}
\def\beqRR{\begin{equation}\label{E:RR}}
\def\beqSS{\begin{equation}\label{E:SS}}
\def\beqTT{\begin{equation}\label{E:TT}}
\def\beqUU{\begin{equation}\label{E:UU}}
\def\beqVV{\begin{equation}\label{E:VV}}
\def\beqWW{\begin{equation}\label{E:WW}}
\def\beqXX{\begin{equation}\label{E:XX}}
\def\beqYY{\begin{equation}\label{E:YY}}
\def\beqZZ{\begin{equation}\label{E:ZZ}}

\def\beqaa{\begin{equation}\label{E:aa}}
\def\beqbb{\begin{equation}\label{E:bb}}
\def\beqcc{\begin{equation}\label{E:cc}}
\def\beqdd{\begin{equation}\label{E:dd}}
\def\beqee{\begin{equation}\label{E:ee}}
\def\beqff{\begin{equation}\label{E:ff}}
\def\beqgg{\begin{equation}\label{E:gg}}
\def\beqhh{\begin{equation}\label{E:hh}}
\def\beqii{\begin{equation}\label{E:ii}}
\def\beqjj{\begin{equation}\label{E:jj}}
\def\beqkk{\begin{equation}\label{E:kk}}
\def\beqll{\begin{equation}\label{E:ll}}
\def\beqmm{\begin{equation}\label{E:mm}}
\def\beqnn{\begin{equation}\label{E:nn}}
\def\beqoo{\begin{equation}\label{E:oo}}
\def\beqpp{\begin{equation}\label{E:pp}}
\def\beqqq{\begin{equation}\label{E:qq}}
\def\beqrr{\begin{equation}\label{E:rr}}
\def\beqss{\begin{equation}\label{E:ss}}
\def\beqtt{\begin{equation}\label{E:tt}}
\def\bequu{\begin{equation}\label{E:uu}}
\def\beqvv{\begin{equation}\label{E:vv}}
\def\beqww{\begin{equation}\label{E:ww}}
\def\beqxx{\begin{equation}\label{E:xx}}
\def\beqyy{\begin{equation}\label{E:yy}}
\def\beqzz{\begin{equation}\label{E:zz}}

\def\beqAAA{\begin{equation}\label{E:AAA}}
\def\beqBBB{\begin{equation}\label{E:BBB}}
\def\beqCCC{\begin{equation}\label{E:CCC}}
\def\beqDDD{\begin{equation}\label{E:DDD}}
\def\beqEEE{\begin{equation}\label{E:EEE}}
\def\beqFFF{\begin{equation}\label{E:FFF}}
\def\beqGGG{\begin{equation}\label{E:GGG}}
\def\beqHHH{\begin{equation}\label{E:HHH}}
\def\beqIII{\begin{equation}\label{E:III}}
\def\beqJJJ{\begin{equation}\label{E:JJJ}}
\def\beqKKK{\begin{equation}\label{E:KKK}}
\def\beqLLL{\begin{equation}\label{E:LLL}}
\def\beqMMM{\begin{equation}\label{E:MMM}}
\def\beqNNN{\begin{equation}\label{E:NNN}}
\def\beqOOO{\begin{equation}\label{E:OOO}}
\def\beqPPP{\begin{equation}\label{E:PPP}}
\def\beqQQQ{\begin{equation}\label{E:QQQ}}
\def\beqRRR{\begin{equation}\label{E:RRR}}
\def\beqSSS{\begin{equation}\label{E:SSS}}
\def\beqTTT{\begin{equation}\label{E:TTT}}
\def\beqUUU{\begin{equation}\label{E:UUU}}
\def\beqVVV{\begin{equation}\label{E:VVV}}
\def\beqWWW{\begin{equation}\label{E:WWW}}
\def\beqXXX{\begin{equation}\label{E:XXX}}
\def\beqYYY{\begin{equation}\label{E:YYY}}
\def\beqZZZ{\begin{equation}\label{E:ZZZ}}

\def\beqaaa{\begin{equation}\label{E:aaa}}
\def\beqbbb{\begin{equation}\label{E:bbb}}
\def\beqccc{\begin{equation}\label{E:ccc}}
\def\beqddd{\begin{equation}\label{E:ddd}}
\def\beqeee{\begin{equation}\label{E:eee}}
\def\beqfff{\begin{equation}\label{E:fff}}
\def\beqggg{\begin{equation}\label{E:ggg}}
\def\beqhhh{\begin{equation}\label{E:hhh}}
\def\beqiii{\begin{equation}\label{E:iii}}
\def\beqjjj{\begin{equation}\label{E:jjj}}
\def\beqkkk{\begin{equation}\label{E:kkk}}
\def\beqlll{\begin{equation}\label{E:lll}}
\def\beqmmm{\begin{equation}\label{E:mmm}}
\def\beqnnn{\begin{equation}\label{E:nnn}}
\def\beqooo{\begin{equation}\label{E:ooo}}
\def\beqppp{\begin{equation}\label{E:ppp}}
\def\beqqqq{\begin{equation}\label{E:qqq}}
\def\beqrrr{\begin{equation}\label{E:rrr}}
\def\beqsss{\begin{equation}\label{E:sss}}
\def\beqttt{\begin{equation}\label{E:ttt}}
\def\bequuu{\begin{equation}\label{E:uuu}}
\def\beqvvv{\begin{equation}\label{E:vvv}}
\def\beqwww{\begin{equation}\label{E:www}}
\def\beqxxx{\begin{equation}\label{E:xxx}}
\def\beqyyy{\begin{equation}\label{E:yyy}}
\def\beqzzz{\begin{equation}\label{E:zzz}}

\def\beqAAAA{\begin{equation}\label{E:AAAA}}
\def\beqBBBB{\begin{equation}\label{E:BBBB}}
\def\beqCCCC{\begin{equation}\label{E:CCCC}}
\def\beqDDDD{\begin{equation}\label{E:DDDD}}
\def\beqEEEE{\begin{equation}\label{E:EEEE}}
\def\beqFFFF{\begin{equation}\label{E:FFFF}}
\def\beqGGGG{\begin{equation}\label{E:GGGG}}
\def\beqHHHH{\begin{equation}\label{E:HHHH}}
\def\beqIIII{\begin{equation}\label{E:IIII}}
\def\beqJJJJ{\begin{equation}\label{E:JJJJ}}
\def\beqKKKK{\begin{equation}\label{E:KKKK}}
\def\beqLLLL{\begin{equation}\label{E:LLLL}}
\def\beqMMMM{\begin{equation}\label{E:MMMM}}
\def\beqNNNN{\begin{equation}\label{E:NNNN}}
\def\beqOOOO{\begin{equation}\label{E:OOOO}}
\def\beqPPPP{\begin{equation}\label{E:PPPP}}
\def\beqQQQQ{\begin{equation}\label{E:QQQQ}}
\def\beqRRRR{\begin{equation}\label{E:RRRR}}
\def\beqSSSS{\begin{equation}\label{E:SSSS}}
\def\beqTTTT{\begin{equation}\label{E:TTTT}}
\def\beqUUUU{\begin{equation}\label{E:UUUU}}
\def\beqVVVV{\begin{equation}\label{E:VVVV}}
\def\beqWWWW{\begin{equation}\label{E:WWWW}}
\def\beqXXXX{\begin{equation}\label{E:XXXX}}
\def\beqYYYY{\begin{equation}\label{E:YYYY}}
\def\beqZZZZ{\begin{equation}\label{E:ZZZZ}}

\def\bal{\begin{aligned}}

\def\eal{\end{aligned}}

\title{Gravitational and electroweak unification by replacing diffeomorphisms with larger group
\input{table}}
\author{Dave Pandres, Jr.$^*$}
\thanks{$^*$North Georgia College and State University, Dahlonega, GA 30597 \\
Emeritus Professor of Mathematics \\
Current Address: 1244 Arbor Rd.,\,\#522, Winston Salem, NC 27104\\
Current E-Mail Address: dpandres@triad.rr.com}
\begin{document}

\begin{abstract}
The covariance group for general relativity, the diffeomorphisms, is replaced by a group of coordinate transformations which contains the diffeomorphisms as a proper subgroup. The larger group is defined by the assumption that all observers will agree whether any given quantity is conserved. Alternatively, and equivalently, it is defined by the assumption that all observers will agree that the general relativistic wave equation describes the propagation of light. Thus, the group replacement is analogous to the replacement of the Lorentz group by the diffeomorphisms that led Einstein from special relativity to general relativity, and is also consistent with the assumption of constant light velocity that led him to special relativity. The enlarged covariance group leads to a non-commutative geometry based not on a manifold, but on a nonlocal space in which paths, rather than points, are the most primitive invariant entities. This yields a theory which unifies the gravitational and electroweak interactions. The theory contains no adjustable parameters, such as those that
are chosen arbitrarily in the standard model.
\end{abstract}


\maketitle

\section{INTRODUCTION}
During the past several decades, we have published nine papers in which we have pursued the suggestion by Einstein  that a unified field theory be constructed by replacing the covariance group for general relativity with a larger group, analogous to the step which led from special relativity to general relativity. Our results unify the gravitational and electroweak interactions.

The purpose of this article is to survey our work in a way that makes it as easy to understand as possible. Thus, we
recall only what is essential and make some changes in notation. In order to optimize readability, we give the detailed calculations justifying certain equations in an Appendix.

\subsection{Suggestions of Dirac and Einstein} Our research has been motivated philosophically, not only by Einstein's [1] suggestion, but also by a previous and more general suggestion by Dirac [2].
\newline\newline
{\bf Dirac:} In the Preface to the First Edition of his famous book {\sl Quantum Mechanics}, Dirac made the prophetic statement:
``The growth of the use of transformation theory, as applied first to relativity and later to the quantum theory, is the essence of the new method in theoretical physics. Further progress lies in the direction of making our equations invariant under wider
\newline
and still wider transformations.''\newline
{\bf Einstein:} In his {\sl Autobiographical Notes}, Einstein made the following statement which is consistent with that of Dirac, but which is more specific: ``A theory can be tested by experience, but there is no way from experience to the setting up of a theory. Equations of such complexity as are the equations of the gravitational field can be found only through the discovery of a logically simple mathematical condition which determines the equations completely or [at least] almost completely. Once one has those sufficiently strong formal conditions, one requires only a little knowledge of facts for the setting up of a theory; in the case of the equations of gravitation it is the four-dimensionality and the
symmetric tensor as expression for the structure of space which, together with the invariance concerning the continuous
transformation-group, determine the equations almost completely.''

``Our problem is that of finding the field equations for the total field. The desired structure must be a generalization of the symmetric tensor. The group must not be any narrower than that of the continuous transformations of co-ordinates. If one introduces a richer structure, then the
group will no longer determine the equations as strongly as in the case of the symmetric tensor as structure. Therefore it would be most beautiful, if one were to succeed in expanding the group once more, analogous to the step which led
from special relativity to general relativity.''
\subsection{The Diffeomorphism Group}In the modern literature, the group that Einstein called the ``continuous transformations of co-ordinates'' is called the ``group of diffeomorphisms.''
For a diffeomorphism between space-time coordinates $x^\alpha$ and $x^{\widetilde\alpha}$, infinitesimal changes in the coordinates are related by $\text d x^{\widetilde\alpha} = x^{\widetilde\alpha}{}\!_\mu \,\text d x^\mu$, where repeated indices are summed over the values $0,1,2,3$. The matrix of the transformation coefficients $x^{\widetilde\alpha}{}\!_\mu$ must be non-singular, and the integrability condition
\begin{equation}\label{E:A}
x^{\widetilde\alpha}{}\!_\mu , _\nu - x^{\widetilde\alpha}{}\!_\nu , _\mu = 0     
\end{equation}
must be satisfied.

 Ordinary partial differentiation is denoted by a comma or, when more convenient, by the symbol
$\partial$. Thus, the partial derivative of $\Phi$ with respect to
$x^\mu$ may be written $\Phi ,_\mu$ or $\partial _\mu \Phi$.
\subsection{Generalization of Einstein's ``Elevator'' Argument}
Initially [3], we merely suggested that the condition expressed by Eq.~(\ref{E:A}) be abandoned, but did not
propose a condition which should replace it. We now recall our motivation for this suggestion. It is based on an
argument which is a generalization of the ``elevator'' argument that led Einstein from special relativity to general relativity.

We begin with the special relativistic equation of motion for a free particle
\begin{equation}\label{E:HHHH}
\dfrac{{\text d}^2 x^i}{\text d s^2}=0 \quad ,
\end{equation}
where $-\text d s^2 = g_{ij}\,\text d x^i \,\text d x^j$, and $g_{ij}=\text{diag}(-1,1,1,1)$. We consider the image-equation of
Eq.~(\ref{E:HHHH}) under a transformation $\text d x^i = h^i{}\!_\mu \,\text d x^\mu$ between coordinate increments
$\text d x^i$ and coordinate increments $\text d x^\mu$,  in which $h^i{}\!_\mu ,_\nu - h^i{}\!_\nu ,_\mu \ne 0.$
Thus, Eq.~(\ref{E:A}), the integrability condition for diffeomorphisms, is not satisfied. The transformation coefficients
$h^i{}\!_\mu $ are an orthonormal tetrad of vectors in terms of which the general relativistic space-time metric is expressed, i.e.,
$g_{\mu\nu}= g_{ij}\, h^i{}\!_\mu \, h^j{}\!_\nu .$

We find (Appendix) that the image-equation is
\begin{equation}\label{E:C}
\frac{\text d^2 x^\alpha}{\text d s^2}+ \Gamma^\alpha{}\!_{\mu\nu} \, \frac{\text d x^\mu}{\text d s} \frac{\text d x^\nu}{\text d s} = -f^\alpha{}\!_\nu \frac{\text d x^\nu}{\text d s} \quad , 
\end{equation}
where $\Gamma^\alpha{}\!_{\mu\nu}$ is the usual Christoffel symbol, and $f_{\mu\nu}=v_i f^i{}\!_{\mu\nu}$, where
$f^i{}\!_{\mu\nu}=h^i{}\!_\nu , _\mu - h^i{}\!_\mu , _\nu$ and $v^i=\dfrac{\text d x^i}{\text d s}$ is the (constant) first integral
of $\dfrac{\text d^2 x^i}{\text d s^2}=0$.  Thus, it appears at first glance that Eq.~(\ref{E:C}) describes a charged particle moving in a gravitational and electromagnetic field.
However, $f_{\mu\nu}$ cannot be interpreted as the electromagnetic field, because the relation
$v^i=\dfrac{\text d x^i}{\text d s}=h^i{}\!_\mu \, \dfrac{\text d x^\mu}{\text d s}$ implies that $v^i$ depends upon $\dfrac{\text d x^\mu}{\text d s}$. Although
$f_{\mu\nu}$ is a linear combination of the $f^i{}\!_{\mu\nu}$ with coefficients $v_i$ which are constant along the
world-line of a particle, it is unsatisfactory that the values of these coefficients should depend upon
$\dfrac{\text d x^\mu}{\text d s}$. This would imply that the electromagnetic field experienced by a particle depends upon the velocity of the particle, in disagreement with experiment.

In two later papers [4,5], we tried to get a satisfactory description of gravitation and electromagnetism alone by investigating a class of theories in which the four antisymmetric tensors $f^i{}\!_{\mu\nu}$ are constant multiples of one another. However, theories in this class suffer from the defect that they are not based on a covariance
group, and thus lack a ``guiding principle.''

The inescapable fact is that for
transformations with $h^i{}\!_\mu , _\nu - h^i{}\!_\nu , _\mu \ne 0$, the right side of Eq.~(\ref{E:C}) contains
four linearly independent antisymmetric tensors, while only one is needed to describe the electromagnetic field.

However, the analysis of this Section does suggest that Eq.~(\ref{E:A}) be abandoned, and that $f^0{}\!_{\mu\nu}$ might somehow describe the electromagnetic field, while $f^i{}\!_{\mu\nu}$ for $i \ne 0$ might somehow describe the short-range interactions.\newline


\subsection{Consequences of Abandoning the Integrability Condition for Diffeomorphisms: Path-Dependence}
If the integrability condition of\linebreak
Eq.~(\ref{E:A}) is not satisfied, it still is possible to ``integrate'' the
relation $\text d x^{\widetilde\alpha}= x^{\widetilde\alpha}{}\!_\mu \, \text d x^\mu$ from the origin $o$ of an $X$ coordinate system to a terminal point $x$ with coordinates $x^\alpha$, and to ``integrate'' the
inverse relation $\text d x^\alpha= x^\alpha{}\!_{\widetilde\mu} \, \text d x^{\widetilde\mu}$ from the origin ${\widetilde o}$ of an ${\widetilde X}$ coordinate system to a terminal point
${\widetilde x}$ with coordinates $x^{\widetilde\alpha}$. (The origins may be regarded as coordinates of observers $\Omega$ and $\widetilde\Omega$ who use the $X$ and ${\widetilde X}$ coordinate systems, respectively.) This gives the results
$
x^{\widetilde\alpha}=\int_o^x x^{\widetilde\alpha}{}\!_\mu \, \text d x^\mu
$
and
$
x^\alpha=\int_{\widetilde o}^{\widetilde x} x^\alpha{}\!_{\widetilde\mu} \,\text d x^{\widetilde\mu} \, .
$
The numerical value of $x^{\widetilde\alpha}$ depends upon the detailed structure of the path $p$ from $o$ to $x$, and the numerical value of $x^\alpha$ depends upon the detailed structure of the path ${\widetilde p}$ from ${\widetilde o}$ to ${\widetilde x}$. There exist many paths $p$ which yield the same value of $x^{\widetilde\alpha}$, and many paths ${\widetilde p}$ which yield the same value of
$x^\alpha$. This means that the correspondence between the points $x$ and ${\widetilde x}$ defined by the coordinates $x^\alpha$ and
$x^{\widetilde\alpha}$ is both one-to-many and many-to-one, rather than one-to-one. However, there does exist a one-to-one correspondence between the paths $p$ and ${\widetilde p}$. Thus, if Eq.~(\ref{E:A}) is abandoned, the space is not a manifold, i.e., a space in which {\sl points} are the most primitive geometrical (invariant) entities. It is instead a nonlocal space in which {\sl paths} are the most primitive geometrical entities.

The fact that paths are our most primitive geometrical entities is interesting because the most intuitive description of quantum theory is Feynman's [6] path-integral formalism, and because a particle localized on a path is a {\sl string.} Thus, it seems possible that if Eq.~(\ref{E:A}) is abandoned, this may lead serendipitously to what Witten [7] has called a much needed ``conceptual, logical framework in which string theory is as natural as general relativity is in its terms.''

Many investigators (see, e.g. Eddington [8]; Penrose, et.~al. [9,10]; Finkelstein, et.~al. [11-13];
Bergmann and Komar [14]; Gambini and Trias [15]) have expressed skepticism that a manifold adequately describes physical space. More recently
and less formally, the same skepticism has been expressed by the Fields medalist Connes [16],  and by Crane [17].
\subsection{\bf {The Simplicity of Working in Path-Space}}  Let $\Phi$ be a path-\linebreak
dependent functional, i.e., a rule that assigns to each path $p$ a real number $\Phi(p)$. Following a method introduced by us [3] and independently by Mandelstam [18], we define derivatives of a path-dependent functional $\Phi(p)$ by giving the path $p$ an extension from its terminus $x$ while holding the rest of the path completely fixed. With this definition, one may work in path-space almost as if it were a manifold. The main difference is that one must carefully preserve the orders of repeated partial derivatives. The following paragraph explains why this is so.

When we wish to emphasize the path-dependent character of a functional $\Phi$ we will use the notation  $\Phi(p)$. However, our functionals include, as a subclass, the usual one-valued functions of $x^\alpha ,$ i.e., functionals                                                                    which are ``path-dependent'' in the trivial sense that they depend only on the terminus $x$ of the path $p$; for them we use the notation $\Phi(x)$. From a path $p$ let two extended paths $p+\Delta p_1$ and $p+\Delta p_2$ be constructed such that the extensions $\Delta p_1$ and $\Delta p_2$ do not completely coincide, but such that the termini of $p+\Delta p_1$ and $p+\Delta p_2$
{\sl do} coincide. The values of $\Phi(p+\Delta p_1)$ and $\Phi(p+\Delta p_2)$ are not generally equal. It follows that
$\partial_\mu \, \partial_\nu \Phi$ and $\partial_\nu\, \partial_\mu\Phi$ are not generally equal. This may be seen by letting $\Delta p_1$ be an
extension along which first only  $x^\mu$ changes and then only $x^\nu$ changes, and letting $\Delta p_2$ be an
extension along which first only  $x^\nu$ changes and then only $x^\mu$ changes. Thus, we see that
$
[{\partial}_\mu , {\partial}_\nu ]\Phi
$
vanishes for functions $\Phi(x),$ but not generally for functionals $\Phi(p)$. The symbol
$
[{\partial}_\mu , {\partial}_\nu ] = {\partial}_\nu \, {\partial}_\mu - {\partial}_\mu \, {\partial}_\nu
$
denotes the usual commutator of partial derivatives.

\section{THE CONSERVATION GROUP}
In Pandres [19], we proposed  that Eq.~(\ref{E:A}) be replaced by the weaker condition
\begin{equation}\label{E:B}
x^\nu{}\!_{\widetilde\alpha} \left(x^{\widetilde\alpha}{}\!_\mu,_\nu - x^{\widetilde\alpha}{}\!_\nu,_\mu\right) = 0  \quad ,
\end{equation}
and proved that this defines a group which contains the diffeomorphisms as a proper subgroup. This proof is given
in Sec.~2.2. The covariance group defined by Eq.~(\ref{E:B}) is our ``guiding principle,'' just as Einstein's requirement
of covariance under the diffeomorphisms was the guiding principle for his theory of gravitation.

\subsection{Justifications for Introducing the Conservation Group}
Pauli [20] emphasized that ``it is absolutely essential to insist that such a fundamental theorem as the covariance law should be derivable from the simplest possible basic assumptions.''
We now recall two (mathematically equivalent) basic assumptions which show that the covariance law based on
Eq.~(\ref{E:B}) satisfies Pauli's requirement.  The first of these assumptions, discussed in Sec.~2.1.1, explains why transformations
that satisfy Eq.~(\ref{E:B}) are called ``conservative.''

\subsubsection{\bf {Agreement Concerning Conserved Quantities}}
The first assumption [19] is that any two observers $\Omega$ and $\widetilde\Omega$, using the coordinates $x^\alpha$ and
$x^{\widetilde\alpha},$ will agree whether or not a given quantity is conserved. A general relativistic conservation law is an expression of the form
$
V^\alpha{}\!_{;\alpha}=0 ,
$
where $V^\alpha$ is a vector density of weight $+1$, and a semicolon denotes covariant differentiation with respect to the  Christoffel symbol. We may however write this as
\begin{equation}\label{E:mmm}
V^\alpha{}\!,_\alpha=0
\end{equation}
because, for a vector density of weight $+1$, the covariant divergence equals the ordinary divergence.

Eq.~(\ref{E:mmm}) is manifestly invariant under the diffeomorphisms. It is surprising, and very important, that Eq.~(\ref{E:mmm}) is also invariant under the larger group defined by Eq.~(\ref{E:B}). To see that this is true, we note that the transformation
law for a vector density of weight $+1$ is \,
\begin{equation}\label{E:D}
V^{\widetilde\alpha}
 =\tfrac{\partial x}{\partial {\widetilde x}} \, x^{\widetilde\alpha}{}\!_\mu V^\mu \quad ,  
\end{equation}
where $\frac{\partial x}{\partial {\widetilde x}}$ is the  determinant of $x^\mu{}\!_{\widetilde\alpha}$.
Upon differentiating Eq.~(\ref{E:D}) with
respect to $x^{\widetilde\alpha}$, after a short calculation (Appendix), we obtain
\begin{equation}\label{E:BBBB}
V^{\widetilde\alpha}{}\!,_{\widetilde\alpha}
=\tfrac{\partial x}{\partial {\widetilde x}} V^\alpha{}\!,{}\!_\alpha +V^\mu \, \tfrac{\partial x}{\partial {\widetilde x}} x^\nu{}\!_{\widetilde\alpha}(x^{\widetilde\alpha}\!_\mu,_\nu-\, x^{\widetilde\alpha}\!_\nu,_\mu).
\end{equation}
Since $\tfrac{\partial x}{\partial {\widetilde x}}\ne 0$,
we see from Eq.~(\ref{E:BBBB}) that, for arbitrary $V^\mu,$ we have both $V^\alpha{}\!,_\alpha=0$ and
$V^{\widetilde\alpha}{}\!,_{\widetilde\alpha}=0$ if and only if the transformation between $x^\alpha$ and $x^{\widetilde\alpha}$ satisfies Eq.~(\ref{E:B}). For this reason, we call transformations that satisfy Eq.~(\ref{E:B}) ``conservative.''
\subsubsection{\bf {Agreement Concerning Equation for propagation of Light}}
The second assumption [21] is that any two observers will agree that the equation of
motion for the propagation of light is the general relativistic wave equation
$
{\dfrac{1}{\sqrt{-g}}} \left(\sqrt{-g}g^{\alpha\nu} {\varPhi}\! ,_\nu \! \right)\! ,_\alpha =0 \, .
$
Thus, our second assumption is that
$
\left(\sqrt{-g}g^{\alpha\nu} {\varPhi}\! ,_\nu \! \right)\! ,_\alpha =0
$
if and only if
$
 \left(\sqrt{-{\widetilde g}}g^{{\widetilde \alpha}{\widetilde \nu}} {\varPhi}\! ,
_{\widetilde \nu} \! \right)\! ,_ {\widetilde \alpha} =0 \, .
$
This second assumption is mathematically equivalent to the first assumption since
$
\sqrt{-g}g^{\alpha\nu}{\varPhi}\! ,_\nu
$
is a vector density of weight $+1$.

The second assumption is especially compelling because it is analogous to the step which led Einstein from special relativity to general relativity, and is also consistent with the assumption of constant light velocity that led him
to special relativity. Thus, the group defined by Eq.~(\ref{E:B}) could be called the ``light-propagation group.''
However, we will retain the name ``conservation group,'' since this name has been used in all prior papers.
\subsection{Proof That Conservative Transformations Form a Group that Contains the Diffeomorphisms as a Proper Subgroup}
We now recall [19] an explicit proof that
transformations which satisfy Eq.~(\ref{E:B}) form a group that contains the diffeomorphisms as a proper subgroup.

First, we note that the identity transformation $x^{\widetilde\alpha} = x^\alpha$ is a
conservative transformation.
Next, we consider the
result of following a  transformation from $x^\alpha$ to $x^{\widetilde\alpha}$ by a
transformation from $x^{\widetilde\alpha} $ to $x^{\widehat\alpha}$. From the relation
\begin{equation}\label{E:F}
x^{\widehat\alpha}\!_\mu = x^{\widehat\alpha}\!_{\widetilde\rho} \, x^{\widetilde\rho}\!_\mu     
\end{equation}
we find (Appendix) that
\begin{equation}\label{E:G}
x^\nu{}\!_{\widehat\alpha} \left(x^{\widehat\alpha}{}\!_\mu,_\nu - x^{\widehat\alpha}{}\!_\nu,_\mu\right)=
x^{\widetilde \sigma}{}\!_{\widehat\alpha} \left(x^{\widehat\alpha}{}\!_{\widetilde \rho} ,_{\widetilde \sigma} -x^{\widehat\alpha}{}\!_{\widetilde \sigma} ,_{\widetilde \rho}\right)x^{\widetilde\rho}\!_\mu + x^\nu{}\!_{\widetilde\alpha} \left(x^{\widetilde\alpha}{}\!_\mu,_\nu - x^{\widetilde\alpha}{}\!_\nu,_\mu\right) \quad  . 
\end{equation}
We see from Eq.~(\ref{E:G}) that if
$
x^\nu{}\!_{\widetilde\alpha} \left(x^{\widetilde\alpha}{}\!_\mu,_\nu - x^{\widetilde\alpha}{}\!_\nu,_\mu\right)
$
and
$
x^{\widetilde \sigma}{}\!_{\widehat\alpha} \left(x^{\widehat\alpha}{}\!_{\widetilde \rho} ,_{\widetilde \sigma} -x^{\widehat\alpha}{}\!_{\widetilde \sigma} ,_{\widetilde \rho}\right)
$
vanish, then
$
x^\nu{}\!_{\widehat\alpha} \left(x^{\widehat\alpha}{}\!_\mu,_\nu - x^{\widehat\alpha}{}\!_\nu,_\mu\right)
$
vanishes. This shows that if the transformations from $x^\alpha$ to $x^{\widetilde\alpha}$
and from $x^{\widetilde\alpha}$ to $x^{\widehat\alpha}$ are conservative,
then the product transformation
from $x^\alpha$ to $x^{\widehat\alpha}$ is conservative.
If we let $x^{\widehat\alpha}=x^\alpha$, we see from
Eq.~(\ref{E:G}) that the inverse of a conservative
transformation is
conservative.
From Eq.~(\ref{E:F}), we see that the product of matrices
$x^{\widetilde\rho}\!_\mu$ and $x^{\widehat\alpha}\!_{\widetilde\rho}$ (which
represent the transformations from
$x^\alpha$ to $x^{\widetilde\alpha}$ and from $x^{\widetilde\alpha}$ to $x^{\widehat\alpha}$, respectively) equals the matrix
$x^{\widehat\alpha}\!_\mu$ (which represents the product transformation from $x^\alpha$ to
$x^{\widehat\alpha}$). It is obvious, and well known, that if products admit a matrix
representation in this sense, then the associative law is satisfied. This
completes the proof that the conservative
transformations form a group.

We note that if Eq.~(\ref{E:A}) is satisfied,
then Eq.~(\ref{E:B}) is satisfied; i.e., the conservation group contains
the diffeomorphisms as a subgroup. Thus, to show that it
contains the diffeomorphisms as a {\sl proper\/} subgroup, we need only
exhibit transformation coefficients which satisfy Eq.~(\ref{E:B}), but do not
satisfy Eq.~(\ref{E:A}). It is easily verified (Appendix) that such transformation coefficients are
\begin{equation}\label{E:H}
\aligned
x^{\widetilde\alpha}\!_\mu &= \delta^\alpha_\mu + \delta^\alpha_0 \delta^2_\mu x^1 \\
x^\nu\!_{\widetilde\alpha} &= \delta^\nu_\alpha - \delta^\nu_0 \delta^2_\alpha x^1
\endaligned
\end{equation}
where $\delta^\alpha_\mu$ is the usual Kronecker delta.
\section{THE CURVATURE VECTOR $C_\mu$}
We define $C_\mu$ by
\begin{equation}\label{E:K}
C_\mu=h_i{}^\nu {} \left(h^i{}\!_\mu ,_\nu-h^i{}\!_\nu ,_\mu\right) \quad . 
\end{equation}
As we have noted in Section~1.1, Einstein [1] stated that introducing a richer structure than the metric would make it desirable to enlarge the covariance group. That the tetrad is such a richer structure may be seen as follows.
If a definite tetrad
$
h^i{}\!_\mu
$
is given, then the metric
$
g_{\mu\nu}
$
is uniquely determined, but the converse is not true. If a definite metric
$
g_{\mu\nu}
$
is given, then the tetrad
$
h^i{}\!_\mu
$
is determined only up to a six-fold parameter of transformations, i.e., the homogeneous Lorentz transformations on the Latin (tetrad)
indices.

We also note that no physics can be done without an observer and that a tetrad represents an observer-frame,
as discussed by Synge [22]. Thus, a theory based on tetrads is especially parsimonious, and hence in accord with
the philosophical principle called Occam's razor. It is also in accord with Dirac's [2] statement that the use of
transformation theory ``is very satisfactory from a philosophical point of view, as implying an increasing recognition of
the part played by the observer in himself introducing the regularities that appear in his observations, and a lack of
arbitrariness in the ways of nature $\dots$ .''

Under a transformation from $x^\alpha$ to $x^{\widetilde\alpha}$, the tetrad transforms according to
$
h^i{}\!_ \mu = h^i{}\!_{{\widetilde\alpha}}\, x^{{\widetilde\alpha}} {}\!_\mu
$
and
$
h_i{}\!^\nu = h_i{}\!^{{\widetilde\alpha}} \, x^ \nu{}\!_{{\widetilde\alpha}}
$.
Thus, we find (Appendix) that the transformation law for
$C_\mu$
is
\begin{equation}\label{E:L}
C_\mu=C_{{\widetilde\alpha}} x^{{\widetilde\alpha}}{}\!_\mu
+ x^\nu{}\!_{{\widetilde\alpha}}\left(x^{{\widetilde\alpha}}{}\!_\mu ,_\nu - x^{{\widetilde\alpha}}{}\!_\nu ,_\mu\right) \quad , 
\end{equation}
where
$
C_{{\widetilde\alpha}}=h_i{}^{{\widetilde\beta}} {} \left(h^i{}\!_{{\widetilde\alpha}} ,_{{\widetilde\beta}}
-(h^i{}\!_{{\widetilde\beta}} ,_{{\widetilde\alpha}}\right) .
$
We see from Eq.~(\ref{E:L}) that $C_\mu$ transforms as a vector if and only if the transformation is conservative.
We now show that
$
C_\mu
$
may be regarded as a {\sl curvature vector,} because $C_\mu$ {\sl vanishes if and only if there exists a conservative transformation from $x^\alpha$ to a special
$x^{\widetilde\alpha}$ in which
$
h^i{}\!_{\widetilde\alpha}
$
is constant.} This may be seen in the following way.  If
$
h^i{}\!_{\widetilde\alpha}
$
is constant then $C_{{\widetilde\alpha}}$ vanishes,
so Eq.~(\ref{E:L}) shows that $C_\mu$ also vanishes if and only if the transformation is conservative. The converse is slightly less direct. Eq.~(\ref{E:L}) shows that any $C_\mu$ and
$C_{{\widetilde\alpha}}$ which both vanish are related by a conservative transformation. Hence, if $C_\mu$ vanishes, then it is related by a conservative transformation to the $C_{{\widetilde\alpha}}$ with
$
h^i{}\!_{{\widetilde\alpha}}
$
constant. Thus, we see that $C_\mu$ may be regarded as a {\sl curvature vector} analogous, for the conservation group, to the Riemann curvature tensor $R^\alpha{}\!_{\beta\mu\nu}$ which, in the conventional (path-independent) theory, vanishes if and only if there exists a diffeomorphism from $x^\alpha$ to a special $x^{\widetilde\alpha}$ in which
$
g_{{\widetilde\mu}\widetilde\nu}
$
is constant.

The definition given in Eq.~(\ref{E:K}) is only the first useful expression for $C_\mu$. There is a second expression, a third expression, and a fourth expression for $C_\mu$, which are sometimes useful

The second expression is obvious, i.e.,
\begin{equation}\label{E:M}
C_\mu= h_i{}^\nu \, f^i{}\!_{\nu\mu}   
\end{equation}
where $f^i{}\!_{\mu\nu}=h^i{}\!_\nu ,_\mu-h^i{}\!_\mu ,_\nu$, as defined in Sec.~1.3.

We see from Eq.~(\ref{E:K}) that
$
C_\mu=h_i{}^\nu {} \left(h^i{}\!_\mu ,_\nu-h^i{}\!_\nu ,_\mu\right)=h_i{}^\nu {} \left(h^i{}\!_{\mu ;\nu}-h^i{}\!_{\nu ;\mu}\right)
=\gamma^\nu{}\!_{\mu\nu}-\gamma^\nu{}\!_{\nu\mu} \, ,
$
where $\gamma_{i\mu\nu}=h_{i\mu;\nu}$ is the Ricci rotation coefficient (Eisenhart~[23]).

[The tetrad is used to convert between Greek (space-time) indices and Latin (tetrad) indices, e.g.,
$
\gamma_{\alpha\mu j} = h^i{}\!_\alpha \, \gamma_{i\mu\nu} \, h_j{}\!^\nu{}
$.
Tetrad indices are raised and lowered by using
$
g^{ij}
$
and
$
g_{ij}
$,
just as space-time indices are raised and lowered by using
$
g^{\mu\nu}
$
and
$
g_{\mu\nu}
$.
Thus, for example
$
\gamma^i{}_{\mu\nu}=g^{ij} \gamma_{j\mu\nu}
$,
and
$
 \gamma^\alpha{}\!_{\mu\nu}=  h_i{}\!^\alpha \gamma^i{}_{\mu\nu}
$.]

Now $\gamma^\nu{}\!_{\nu\mu}$ vanishes, because, as Eisenhart noted, the rotation coefficients are antisymmetric in their first two indices. Thus, from
$
C_\mu=\gamma^\nu{}\!_{\mu\nu}-\gamma^\nu{}\!_{\nu\mu} \, ,
$
we have our third expression
\begin{equation}\label{E:N}
C_\mu=\gamma^\nu{}\!_{\mu\nu} \quad .  
\end{equation}
The fourth expression will be given in Sec.~5.3.

The antisymmetry of $\gamma_{\mu\nu i}$ in $\mu$ and $\nu$  may be used (Appendix) to obtain an expression for
$
\gamma_{\mu\nu i}
$
in terms of
$
f_{i\mu\nu}
$.
The result is
\begin{equation}\label{E:i}
\gamma_{\mu\nu i} = \tfrac{1}{2} \left(f_{i\mu\nu} - f_{\mu\nu i} - f_{\nu i\mu}\right) 
\; .
\end{equation}

\section{AN IDENTITY FOR THE EINSTEIN TENSOR}
Before recalling our theory which is covariant under the conservation group, it is convenient to obtain an identity
for the Einstein tensor in terms of the Ricci rotation coefficients.

The Riemann curvature tensor is defined by
$$
R^\alpha{}\!_{\beta\mu\nu} = h_i{}\!^\alpha \left(h^i{}_{\beta ;\mu ;\nu} - h^i{}_{\beta ;\nu ;\mu}\right).
$$
It is easily seen from the relation
$
\left(h_i{}\!^\alpha h^i{}_{\beta ;\mu}\right)_{;\nu}=h_i{}\!^\alpha {}\, h^i{}\!_{\beta;\mu;\nu} + h_i{}\!^\alpha {}\!_{;\nu}\, h^i{}\!_{\beta;\mu}
$
that the expression for
$
R^\alpha{}\!_{\beta\mu\nu}
$
may be manipulated into the form
\begin{equation}\label{E:Q}
R^\alpha{}\!_{\beta\mu\nu} = \gamma^\alpha{}\!_{\beta\mu;\nu} - \gamma^\alpha{}\!_{\beta\nu;\mu}+\gamma^\alpha{}\!_{\sigma\nu}\,\gamma^\sigma{}\!_{\beta\mu}
                                                             -\gamma^\alpha{}\!_{\sigma\mu}\,\gamma^\sigma{}\!_{\beta\nu}   
\end{equation}
which is the well known Ricci identity for the curvature tensor (see, e.g., Taub [24])  in terms of the Ricci rotation coefficients.

The Einstein tensor is defined as usual by
$
G_{\mu\nu}=R_{\mu\nu}-\tfrac{1}{2}g_{\mu\nu}R \, ,
$
where
$
R_{\mu\nu}=R^\alpha{}\!_{\mu\alpha\nu}
$
is the usual Ricci tensor, and
$
R=R^\alpha{}\!_\alpha
$
is the usual Ricci scalar.
From Eq.~(\ref{E:Q}) and Eq.~(\ref{E:N}), it follows easily that identities for the Ricci tensor and the Ricci scalar
are
\begin{equation}\label{E:R}
R_{\mu\nu} = C_{\mu;\nu}-C_\alpha\,\gamma^\alpha{}\!_{\mu\nu} +\gamma_\mu{}\!^\alpha{}\!_{\nu;\alpha}   +\gamma^\alpha{}\!_{\sigma\nu}\,\gamma^\sigma{}\!_{\mu\alpha} \quad , 
\end{equation}
and
\begin{equation}\label{E:S}
R=2C^\alpha{}\!_{;\alpha} + C^\alpha {}\, C_\alpha -\gamma^{\alpha\sigma\beta}{}\, \gamma_{\beta\sigma\alpha} \quad .  
\end{equation}
Thus, our identity for the Einstein tensor is
\begin{equation}\label{E:T}
\begin{aligned}
G_{\mu\nu} =& \,C_{\mu;\nu}-C_\alpha \gamma^\alpha{}\!_{\mu\nu}
     -g_{\mu\nu} C^\alpha{}\!_{;\alpha}
   -\tfrac{1}{2} g_{\mu\nu} C^\alpha C_\alpha \\
          & + \gamma_\mu{}\!^\alpha{}\!_{\nu;\alpha} + \gamma^\alpha{}\!_{\sigma\nu} \gamma^\sigma{}\!_{\mu\alpha}
              + \tfrac{1}{2} g_{\mu\nu} \gamma^{\alpha\sigma\beta} \gamma_{\beta\sigma\alpha} \quad . 
\end{aligned}
\end{equation}

\section{THEORY COVARIANT UNDER THE CONSERVATION GROUP}
\subsection{The Lagrangian}
In [19], we chose
$
g^{\mu\nu} C_\mu C_\nu
$
as our Lagrangian because it is the only invariant that can be formed from the curvature vector $C_\mu$ by contraction.
Thus, we used the variational principle
\begin{equation}\label{E:nnn}
\delta \int g^{\mu\nu} C_\mu C_\nu \sqrt{-g}\, d^4\! x = 0 \quad ,  
\end{equation}
where the tetrad $h_j{}^\nu$ is varied.

A tedious, but straightforward, calculation (Appendix) shows that this
variational principle leads to the field equations
\begin{equation}\label{E:g}
C_{\mu;\nu}-C_\alpha \gamma^\alpha{}\!_{\mu\nu} -g_{\mu\nu} C^\alpha{}\!_{;\alpha}- \tfrac{1}{2} g_{\mu\nu} C^\alpha C_\alpha = 0 \; .
\end{equation}

Notice that Eq.~(\ref{E:g}) just states that the first line in Eq.~(\ref{E:T}), the identity for the Einstein tensor, vanishes.
Thus, we may write our field equations in the form
\begin{equation}\label{E:h}
G_{\mu\nu} =\gamma_\mu{}\!^\alpha{}\!_{\nu;\alpha} + \gamma^\alpha{}\!_{\sigma\nu} \gamma^\sigma{}\!_{\mu\alpha}
              + \tfrac{1}{2} g_{\mu\nu} \gamma^{\alpha\sigma\beta} \gamma_{\beta\sigma\alpha} \quad .    
\end{equation}
The variational calculation in the Appendix also shows that Noether's theorem yields the six currents
$$
{\frak J}_{ij}{}\!^\nu = C_\alpha(h_i{}\!^\alpha \, h_j{}\!^\nu - h_j{}\!^\alpha \, h_i{}\!^\nu)
$$
which satisfy the conservation laws
$
{\frak J}_{ij}{}\!^\nu{}\!_{:\nu}=0 \, .
$
Indeed, the antisymmetric part of our field equations, Eq.~(\ref{E:g}), just states that
$
{\frak J}_{ij}{}\!^\nu{}\!_{:\nu}=0 \, .
$
It is, however, the symmetric part of the field equations which will display the unification of the gravitational
and electroweak fields.
\subsection{Schr\"odinger's Interpretation of the Einstein Equations}
The Einstein equations of general relativity
$G_{\mu\nu}=T_{\mu\nu}$
relate a geometrical object, the Einstein tensor
$G_{\mu\nu}$ on the left, to a physical object, the stress-energy tensor $T_{\mu\nu}$ on the right. Einstein felt that this was undesirable and that the use of $T_{\mu\nu}$ in this way was at best a provisional solution.

Schr\"odinger [25] proposed an attractive general solution to this problem. He suggested that the Einstein equations should be regarded not as differential equations to be solved for $g_{\mu\nu}$, with $T_{\mu\nu}$ given, but rather as the definition of the stress-energy tensor: ``I would rather that you did not regard these equations as field equations, but
rather as a definition of $T_{ik}$, the matter tensor.''

In our theory, however, this is not quite satisfactory,
because the right-hand side of our Eq.~(\ref{E:h}) for $G_{\mu\nu}$ in not generally symmetric in $\mu$ and $\nu.$
This is not a contradiction because, as we have stated in Sec.~1.5, when quantities are path-dependent,
one must carefully preserve the orders of repeated partial derivatives. Taking this into account, one finds
(Appendix) that
\begin{equation}\label{E:AAAA}
G_{\mu\nu}-G_{\nu\mu} = h_i{}\!^\alpha \left(\left[\partial_\nu , \partial_\alpha\right] h^i{}\!_\mu + \left[\partial_\alpha , \partial_\mu\right] h^i{}\!_\nu
                     + [\partial_\mu ,\partial_\nu]h^i{}\!_\alpha \right)
                    \quad .
\end{equation}
Thus, $G_{\mu\nu}$ is not generally symmetric if the tetrad is path-dependent. (We suspect
that any tetrad which satisfies the field equations admits a conservative transformation to a coordinate system in which
it is path-independent, but we have not succeeded in proving this.)

The stress-energy tensor $T_{\mu\nu}$, however, must be symmetric in order to have the needed physical properties. We therefore adopt a slight modification of Schr\"odinger's proposal, in which we define $T_{\mu\nu}=G_{\underline{\mu\nu}}$, where
$G_{\underline{\mu\nu}}=\tfrac{1}{2}(G_{\mu\nu}+G_{\nu\mu})$ is the symmetric part of $G_{\mu\nu}.$ Thus, from Eq.~(\ref{E:h}), we
obtain
\begin{equation}\label{E:yyy}
T_{\mu\nu} =\tfrac{1}{2}(\gamma_\mu{}\!^\alpha{}\!_{\nu;\alpha}+\gamma_\nu{}\!^\alpha{}\!_{\mu;\alpha}   +\gamma^\alpha{}\!_{\sigma\nu} \gamma^\sigma{}\!_{\mu\alpha}
          + \gamma^\alpha{}\!_{\sigma\mu} \gamma^\sigma{}\!_{\nu\alpha}+  g_{\mu\nu} \gamma^{\alpha\sigma\beta} \gamma_{\beta\sigma\alpha}) \quad .    
\end{equation}

\subsection{Physical Interpretation for the Stress-Energy Tensor}
In order to manipulate Eq.~(\ref{E:yyy}) into a form such that the physical interpretation of $T_{\mu\nu}$ is clear, we
have used [26] the permutation group decomposition of the Ricci rotation coefficient $\gamma_{\mu\nu i}.$

The permutation group on three indices has six group elements. One element is the
identity. The other five group elements are ``cycles'' such as $(\mu\nu i)$,
which has the effect of replacing $\mu$ with $\nu$, $\nu$ with $i$, and $i$ with
$\mu$. These five group elements are $(\mu\nu)$, $(\nu i)$, $(i\mu)$, $(\mu i\nu)$,
and $(\mu\nu i)$.

We have
\begin{equation}\label{E:l}
\gamma_{\mu\nu i}=A_{\mu\nu i}+M_{\mu\nu i} \;  ,      
\end{equation}
where $A_{\mu\nu i}$ is the totally antisymmetric part, and $M_{\mu\nu i}$ is the ``mixed symmetry'' part of $\gamma_{\mu\nu i}.$
(The totally symmetric part vanishes because $\gamma_{\mu\nu i}$ is antisymmetric in its first two indices.)

The totally antisymmetric part is
\begin{equation}\label{E:j}
A_{\mu\nu i}=\tfrac{1}{3} \left(\gamma_{\mu\nu i} +\gamma_{\nu i\mu}+\gamma_{i\mu\nu}\right) 
\;  .
\end{equation}
Hence, the mixed symmetry part is
$
M_{\mu\nu i}=\gamma_{\mu\nu i}-A_{\mu\nu i \, .}
$
This gives
\begin{equation}\label{E:k}
M_{\mu\nu i}=\tfrac{1}{3}\left(2\gamma_{\mu\nu i} -\gamma_{\nu i\mu}-\gamma_{i\mu\nu}\right)=\tfrac{1}{3}\left(2\gamma_{\mu\nu i} +\gamma_{i\nu\mu}-\gamma_{i\mu\nu}\right) \; ,
\end{equation}
which is antisymmetric in $\mu$ and $\nu$. We shall see (as discussed in [27]) that $M_{\mu\nu i}$ describes the electroweak field.

We see from Eq.~(\ref{E:N}) and Eq.~(\ref{E:l}) that
\begin{equation}\label{E:m}
C_\mu=M^\nu{}\!_{\mu\nu} \;    
\end{equation}
which is our fourth expression for $C_\mu$.

We also need a vector $A^\mu$ which is essentially the dual of $A_{\alpha\beta\sigma},$ i.e.,
\begin{equation}\label{E:x}
A^\mu= \tfrac{1}{3!}(-g)^{-1/2}e^{\mu\alpha\beta\sigma}A_{\alpha\beta\sigma} \quad .
\end{equation}
where $e^{\mu\alpha\beta\sigma}$ is the usual Levi-Civita tensor density of weight $+1$.

If we substitute Eq.~(\ref{E:l}) into Eq.~(\ref{E:yyy}), a tedious calculation (Appendix) gives
\begin{equation}\label{E:EE}
\aligned
T_{\mu\nu}=
         & \tfrac{1}{2}\left( J_{\mu i} h^i{}\!_\nu + J_{\nu i} h^i{}\!_\mu\right)
         -( M^\alpha{}\!_{\mu i} M_{\alpha\nu}{}\!^i - \tfrac{1}{4} g_{\mu\nu} M^{\alpha\sigma i} M_{\alpha\sigma i})\\
          &+2A_\mu A_\nu + g_{\mu\nu} A^\alpha A_\alpha
\quad ,
\endaligned
\end{equation}
where the
$
J_{\mu i}=M_\mu{}\!^\alpha{}\!_{i;\alpha}
$
are the currents which are the source of the electroweak field
$
M_{\mu\nu i}
$. (If the tetrad is path-dependent, this could explain the experimentally observed electroweak parity breaking, and we note that in the Weinberg-Salam theory this parity breaking is ``put in by hand.''

The quantity
\begin{equation}\label{E:FF}
M_{\mu\nu}= M^\alpha{}\!_{\mu i} M_{\alpha\nu}{}\!^i - \tfrac{1}{4} g_{\mu\nu} M^{\alpha\sigma i} M_{\alpha\sigma i}
\end{equation}
has the formal structure of the stress-energy tensor for the Salam [28] - Weinberg [29] electroweak theory, and
Eq.~(\ref{E:EE})
contains the terms $2A_\mu A_\nu + g_{\mu\nu} A^\alpha A_\alpha$ that correspond to the presence of mass.

It is important to note
that the quantity $M_{\mu\nu}{}\!^i$ appears in our equations because the diffeomorphisms have been replaced by the larger
conservation group. There is no need to introduce $M_{\mu\nu}{}\!^i$ as a ``compensating field'' corresponding to
a $U(1)\, \times SU(2)$ gauge group, as Yang and Mills [30] did in their original paper on gauge theory, motivated by Heisenberg's idea that if the electromagnetic field were ``turned off'' it would be impossible to distinguish between a proton and a neutron. There is also no need to depend upon spontaneous symmetry breaking to bring in Higgs bosons
and mass. Moreover, it is clear that Pauli's requirement that the covariance law be derived from the simplest possible basic assumptions is met less satisfactorily by the assumptions that justify gauge theory than by either our assumption that all observers will agree whether a given quantity is conserved, or our assumption that all observers will agree that the propagation of light is described by the general relativistic wave equation.

Equation~(\ref{E:FF}) is manifestly invariant under homogeneous Lorentz transformations on the Latin indices. However, these transformations mimic those of the $U(1)\times SU(2)$ gauge group. This may be seen in the following way. We find from
Eq.~(\ref{E:i}) and Eq.~(\ref{E:k}) that
\begin{equation}\label{E:GG}
M_{\mu\nu i}=\tfrac{1}{3} \left(2f_{i\mu\nu} - f_{\mu\nu i} - f_{\nu i\mu}\right)\quad ,
\end{equation}
which may be written
\begin{equation}\label{E:HH}
M_{\mu\nu i}=\tfrac{1}{3}\left(2 \delta^n_i  \delta^\alpha_\mu \delta^\sigma_\nu
                      - h^n{}\!_\mu \delta^\alpha_\nu h_i{}\!^\sigma
                      - h^n{}\!_\nu  h_i{}\!^\alpha  \delta^\sigma_\mu\right)f_{n\alpha\sigma} \quad .
\end{equation}
It is interesting, and perhaps important, to notice that Eq.~(\ref{E:HH}) may be rewritten into the form
\begin{equation}\label{E:II}
M_{\mu\nu i}=\tfrac{1}{3}\left(2 \delta^n_i \delta^\alpha_\mu \delta^\sigma_\nu
                      - h^n{}\!_\mu \delta^\alpha_\nu h_i{}\!^\sigma
                      - h^n{}\!_\nu  h_i{}\!^\alpha  \delta^\sigma_\mu\right)
                        \frak F_{n\alpha\sigma} \quad  ,
\end{equation}
where
\begin{equation}\label{E:JJ}
\frak F_{n\alpha\sigma} = f_{n\alpha\sigma} + e_{0njk} h^j{}\!_\alpha h^k{}\!_\sigma
\end{equation}
is the usual field strength (see, e.g., Nakahara [31]) for a $U(1)\times SU(2)$ gauge field. In rewriting Eq.~(\ref{E:HH}) as Eq.~(\ref{E:II}), we have used the easily verifiable fact that
$$
\left(2 \delta^n_i \delta^\alpha_\mu \delta^\sigma_\nu
                      - h^n{}\!_\mu \delta^\alpha_\nu h_i{}\!^\sigma
                      - h^n{}\!_\nu  h_i{}\!^\alpha  \delta^\sigma_\mu\right)
                         e_{0njk} h^j{}\!_\alpha h^k{}\!_\sigma = 0 \quad .
$$
From Eq.~(\ref{E:II}), we see that in the expression, Eq.~(\ref{E:HH}), for $M_{\mu\nu i}$, the
curl $f_{n\alpha\sigma}$ may simply be replaced by the gauge field strength $\frak F_{n\alpha\sigma}.$
In our theory, the quantity
$
\frak F_{n \alpha\sigma}
$
does not directly describe the electroweak field. It is, however, the
fundamental quantity which is essential for the construction of that field.
The
$
\frak F_{n\alpha\sigma}
$
in Eq.~(\ref{E:II})
may be viewed as a field with ``bare'' or massless quanta, which
is ``clothed'' by the factor
$
\tfrac{1}{3}\left(2 \delta^n_i \delta^\alpha_\mu \delta^\sigma_\nu
                      - h^n{}\!_\mu \delta^\alpha_\nu h_i{}\!^\sigma
                      - h^n{}\!_\nu  h_i{}\!^\alpha  \delta^\sigma_\mu\right)\, .
$  It is $M_{\mu\nu i}$ that we
identify as the physical electroweak field, and which
appears in the appropriate way in the stress-energy tensor of the Einstein equations.
For this identification to be valid,
the quantity
$
M_{\mu\nu 0}=\tfrac{1}{3} \left(2f_{0\mu\nu}
- f_{\mu\nu 0} - f_{\nu 0\mu}\right)
$
must describe the electromagnetic field; hence, it must be the curl of a vector.
The presence of the terms $- f_{\mu\nu 0} - f_{\nu 0\mu}$ may cause one to ask how $M_{\mu\nu 0}$ can be identified as the electromagnetic field.

Our answer is this: The orthodox physical interpretation, which we adopt, is that $h^i{}\!_\mu$
describes an observer-frame.  Now, \sl{if $h^i{}\!_\mu$ describes a freely-falling,
nonrotating observer frame, our expression for $M_{\mu\nu 0}$ reduces to
$M_{\mu\nu 0}=\tfrac{1}{3} f_{0\mu\nu}$.} This may be seen as follows.
The condition for a freely-falling, nonrotating frame (Synge [22]) is
$h_{i\nu;\alpha} h_0{}\!^\alpha =0$. In terms of the Ricci rotation
coefficients, the condition is
$
\gamma_{\mu\nu 0}=0 .
$ \;
From this and Eq.~(\ref{E:k}), we see that for an $h^i{}\!_\mu$ which describes a
freely falling, nonrotating observer frame,
$
M_{\mu\nu 0}=\tfrac{1}{3}\left(\gamma_{0\nu\mu} -\gamma_{0\mu\nu}\right)
          =\tfrac{1}{3}\left(h_{0\nu;\mu} - h_{0\mu;\nu}\right)
          =\tfrac{1}{3}\left(h_{0\nu},\!_\mu - h_{0\mu},\!_\nu\right)
          =\tfrac{1}{3} f_{0\mu\nu}
$.

\subsection{Further Insight Into the Lagrangian} In retrospect, we can now see intuitively why the Lagrangian
$g^{\mu\nu}C_\mu C_\nu = C^\alpha C_\alpha$ yields a theory which unifies the gravitational and electroweak interactions, and includes a term that brings in mass.

We see from Eq.~(\ref{E:S}) that
\begin{equation}\label{E:MM}
                        C^\alpha {}\, C_\alpha =R+\gamma^{\mu i\nu}{}\, \gamma_{\mu\nu i}-2C^\alpha{}\!_{;\alpha}        \quad .  
\end{equation}
The first term on the right side of Eq.~(\ref{E:MM}) is the Ricci scalar, which is the Lagrangian for gravitation. The last term is a covariant divergence, which contributes nothing to the field equations. After a bit of work (Appendix), we obtain
$
\gamma^{\mu i\nu} \gamma_{\mu\nu i} = \tfrac{1}{2} M^{\mu\nu i} M_{\mu\nu i} + 6 A^\alpha  A_\alpha  \,  ,
$
and our expression for the Lagrangian becomes
\begin{equation}\label{E:NN}
C^\alpha {}\, C_\alpha =R+\tfrac{1}{2} M^{\mu\nu i} M_{\mu\nu i} + 6 A^\alpha  A_\alpha -2C^\alpha{}\!_{;\alpha}  \quad .
\end{equation}
The term $M^{\mu\nu i} M_{\mu\nu i}$ is in the form of the usual electroweak Lagrangian, and the $A^\alpha  A_\alpha$ term has precisely the form that is needed (see, e.g., Moriyasu [32]) for the introduction of mass.

\subsection{Concerning Quantization} As many investigators agree, a major reason for the difficulty (perhaps
impossibility) in quantizing general relativity is that the Lagrangian $R$ contains higher than first derivatives
of the metric. By contrast, the Lagrangian $C^\alpha C_a$ contains no higher than first derivatives of the tetrad.
Also, $C^\alpha C_a$ is only quadratic in these first derivatives. We are therefore optimistic that quantization
of the theory with Lagrangian $C^\alpha C_a$ may be much more straightforward than the theory with Lagrangian $R.$
Previously [33], we have used the Dirac-Bergmann algorithm (Bergmann and Goldberg [34], Dirac [35],
Sundermeyer [36], Weinberg [37]) for constrained dynamics to find the consistent Hamiltonian for this theory.

\section{CONCLUSIONS}
Einstein introduced the diffeomorphisms in order to accommodate observers who are accelerated with respect to one another. But, is there a need for the inclusion of some still more general class of observers? In this Section, we recall a somewhat speculative argument [38] that there is such a need.

General relativity makes use of a classical observer who can observe the motion of a physical system without disturbing the
system. This violates the fundamental principles of quantum theory. Most discussions of observation in quantum theory make use of a ``macroscopic'' classical observer - one who can ``stand outside'' a quantum mechanical system and act upon the system without being acted upon by the system. This is unsatisfactory, because there exist no observers who are
\sl{infinitely large} by comparison with quantum mechanical systems. One solution of this problem was given by Everett [39]. He considers a quantum observer's memory to have quantum states that are correlated with the states of what he has observed. Each observer can then consider himself a macroscopic observer (since his different states are independent)
and still treat other observers as part of his quantum mechanical universe. Each observer can also assign coordinates to
events in his universe. The uncertainty principle does not limit the precision with which he can do this, because the
four operators which represent his coordinates commute. Let $\Omega$ and $\widetilde\Omega$ be two observers, each of whom considers himself
a macroscopic observer, while treating the other as part of his quantum mechanical universe. Let $x^\alpha$ and $x^{{\widetilde\alpha}}$
be coordinates which are assigned by $\Omega$ and $\widetilde\Omega$, respectively, to events in their universes. The assumption that physical space-time is a differentiable manifold rests squarely upon the assertion that it is possible to establish a one-to-one correspondence between $x^\alpha$ and $x^{{\widetilde\alpha}}$, at least in coordinate patches. Einstein challenged the validity of Newton's absolute time on the ground that no operational method had been, or could be, given for its measurement.
In this spirit, we challenge the validity of space-time being a differentiable manifold, on the ground that no operational measurement has been, or can be, given for establishing a one-to-one correspondence between the coordinates $x^\alpha$ and
$x^{{\widetilde\alpha}}$. Our two observers are free to exchange information so that, for example, $\Omega$ can possess a complete description of the procedure which $\widetilde\Omega$ uses in assigning coordinates to events. If $\Omega$ could also state with certainty
(as in general relativity) that $\widetilde\Omega$'s world-line is a particular path $p$, then he could write an expression uniquely
specifying $\widetilde\Omega$'s coordinates $x^{{\widetilde\alpha}}$ in terms of his own coordinates $x^\alpha$. Thus $x^{{\widetilde\alpha}}$ could be regarded as a functional of the known path $p$ with terminus $x$ having the coordinates $x^\alpha$, i.e.,
$
x^{{\widetilde\alpha}}=x^{{\widetilde\alpha}}(x) \, .
$
However, the observer $\Omega$ has described
$\widetilde\Omega$'s world-line as completely as nature permits when he states that \sl{all} paths occur with equal probability amplitude,
in the sense that the probability amplitude for a path $p$ is
$
Ne^{iL(p)/\hbar}\,
$
where $L$ is $\Omega$'s Lagrangian for $\widetilde\Omega$, $N$ is a normalization factor (the same for all paths), and $\hbar$ is the usual
quantum of angular momentum. Therefore, $\Omega$ can only state that the probability amplitude for $\widetilde\Omega$'s coordinate number
$x^{{\widetilde\alpha}}$ corresponding to his coordinate number $x^\alpha$ is
\begin{equation}\label{E:LL}
\Psi(x^\alpha ,x^{{\widetilde\alpha}}) = \sum_p Ne^{iL(p)/\hbar}
\end{equation}
where
$
\sum_p
$
denotes the democratic sum with equal weight of contributions due to all paths $p$ with terminus $x^\alpha$ such that
the image-paths ${\widetilde p}$ have terminus $x^{{\widetilde\alpha}}$. As the sizes of observers $\Omega$ and $\widetilde\Omega$ increase without limit, we find that the competing alternatives in equation (\ref{E:LL}) interfere destructively on all but the classically allowed path. Thus
Eq.~(\ref{E:LL}) goes over to the transformation $x^{{\widetilde\alpha}}=x^{{\widetilde\alpha}}(x),$ with the inverse transformation $x^{\alpha}=x^\alpha ({\widetilde x}),$ in the macroscopic limit.

The above interpretation that conservative transformations relate the measurements of ``quantum observers''
is attractive, but is not essential for our unification of the gravitational and electroweak interactions.

We have obtained our unification by following Einstein's [1] general suggestion that the diffeomorphisms be replaced by
some larger group, i.e., some appropriate covariance group that contains the diffeomorphisms as a proper subgroup. We have obtained such an appropriate group, the conservation group, by assuming that all observers will agree whether or not any given quantity is conserved. Field equations have been obtained from a variational principle with a Lagrangian invariant under the conservation group. (Our Lagrangian is analogous to the conventional Lagrangian $R$ invariant under the diffeomorphisms, used in deriving Einstein's equations for gravitation alone.) Our field equations unify the gravitational and electroweak interactions. The gravitational field is described by the metric, as in general relativity. The electroweak field is described by the ``mixed symmetry'' part of the Ricci rotation coefficients under the permutation group on three indices. Our theory has a stress-energy tensor which agrees with that of the Salam-Weinberg
electroweak theory, and contains terms that correspond to the presence of mass.
\begin{center}
APPENDIX
\end{center}

In this Appendix, we give the detailed computation justifying certain equations in order to optimize readability.\newline

\paragraph{\bf{Equations (\ref{E:HHHH}) and (\ref{E:C})}} From $\text d x^i = h^i{}\!_\mu \, \text d x^\mu$, we have
$
\frac{\text d x^i}{\text d s} = h^i{}\!_\mu \, \frac{\text d x^\mu}{\text d s},
$
so
$$
\frac{\text d^2 x^i}{\text d s^2}= h^i{}\!_\mu \, \frac{\text d^2x^\mu}{\text d s^2}+ \left(\frac{\text d}{\text d s} h^i{}\!_\mu\right) \frac{\text d x^\mu}{\text d s}
                    = h^i{}\!_\mu \, \frac{\text d^2x^\mu}{\text d s^2}+  h^i{}\!_\mu ,_\nu \frac{\text d x^\mu}{\text d s} \frac{\text dx^\nu}{\text d s} \; .
$$
Thus, since
$
\dfrac{\text d^2 x^i}{\text d s^2}=0 \, ,
$
multiplication by $h_i{}\!^\alpha$ gives
\begin{equation}\label{E:IIII}
\frac{\text d^2x^\alpha}{\text d s^2}+ h_i{}\!^\alpha h^i{}\!_\mu ,_\nu \frac{\text d x^\mu}{\text ds }
\dfrac{\text d x^\nu}{\text d s}=0 \, .
\end{equation}
We need an alternative expression for $h_i{}\!^\alpha h^i{}\!_\mu ,_\nu.$
The Christoffel symbol is
\begin{equation}\label{E:JJJJ}
\Gamma^\alpha{}\!_{\mu\nu}=\tfrac{1}{2}g^{\alpha\beta}(g_{\beta\mu} ,_\nu + g_{\beta\nu} ,_\mu - g_{\mu\nu} ,_\beta) \quad .
\end{equation}
From
$
g_{\mu\nu}=g_{ij}h^i{}\!_\mu h^j{}\!_\nu \, ,
$
we get
$
g_{\mu\nu},_\beta =g_{ij}h^i{}\!_\mu ,_\beta h^j{}\!_\nu + g_{ij}h^i{}\!_\mu h^j{}\!_\nu ,_\beta \, ,
$
so
$$
\aligned
g_{\beta\mu},_\nu+g_{\beta\nu},_\mu-g_{\mu\nu},_\beta&=g_{ij} (h^i{}\!_\mu h^j{}\!_\beta ,_\nu + h^i{}\!_\beta h^j{}\!_\mu ,_\nu
                                                + h^i{}\!_\nu h^j{}\!_\beta ,_\mu + h^i{}\!_\beta h^j{}\!_\nu,_\mu
                                                 -h^i{}\!_\nu h^j{}\!_\mu ,_\beta  - h^i{}\!_\mu h^j{}\!_\nu ,_\beta)\\
                                          &=2g_{ij}h^i{}\!_\beta h^j{}\!_\mu,_\nu+g_{ij} (h^i{}\!_\beta f^j{}\!_{\mu\nu}
                                             - h^i{}\!_\mu f^j{}\!_{\beta\nu}
                                             - h^i{}\!_\nu f^j{}\!_{\beta\mu})  \; .
\endaligned
$$
If we multiply the above equation by
$
\frac{1}{2}g^{\alpha\beta}
$
and use Eq.~(\ref{E:JJJJ}), we get
$$
 h_i{}\!^\alpha h^i{}\!_\mu,_\nu=\Gamma^\alpha{}\!_{\mu\nu}- \frac{1}{2}g^{\alpha\beta}g_{ij} (h^i{}\!_\beta f^j{}\!_{\mu\nu}
                                             - h^i{}\!_\mu f^j{}\!_{\beta\nu}
                                             - h^i{}\!_\nu f^j{}\!_{\beta\mu}) \quad .
$$
We multiply this equation by
$
\dfrac{\text d x^\mu}{\text d s} \dfrac{\text d x^\nu}{\text d s} \, ,
$
to get
$$
h_i{}\!^\alpha h^i{}\!_\mu,_\nu \dfrac{\text d x^\mu}{\text d s} \dfrac{\text d x^\nu}{\text d s}=\Gamma^\alpha{}\!_{\mu\nu}\dfrac{\text d x^\mu}{\text d s} \dfrac{\text d x^\nu}{\text d s}
+ g^{\alpha\beta}g_{ij} h^i{}\!_\mu f^j{}\!_{\beta\nu}\dfrac{\text d x^\mu}{\text d s} \dfrac{\text d x^\nu}{\text d s} \quad .
$$
We use this expression for $h_i{}\!^\alpha h^i{}\!_\mu,_\nu \dfrac{\text d x^\mu}{\text d s} \dfrac{\text d x^\nu}
{\text d s}$ in
Eq.~(\ref{E:IIII}) to obtain
\begin{equation}\label{E:LLLL}
\frac{\text d^2x^\alpha}{\text d s^2}+\Gamma^\alpha{}\!_{\mu\nu}\dfrac{\text d x^\mu}{\text d s} \dfrac{\text d x^\nu}
{\text d s}=- g^{\alpha\beta}g_{ij} h^i{}\!_\mu \dfrac{\text d x^\mu}{\text d s} f^j{}\!_{\beta\nu} \dfrac{\text d x^\nu}
{\text d s} \quad .
\end{equation}
Notice that
$
h^i{}\!_\mu \dfrac{\text d x^\mu}{\text d s}= \dfrac{\text d x^i}{\text d s} \, ,
$
which is the (constant) first integral of the free particle equation
$
\dfrac{\text d^2 x^i}{\text d s^2}=0 \, .
$
We denote $h^i{}\!_\mu \dfrac{\text d x^\mu}{\text d s}=\dfrac{\text dx^i}{\text d s}$ by $v^i ,$ and see that Eq.~(\ref{E:LLLL}) becomes
$$
\frac{\text d^2x^\alpha}{\text d s^2}+\Gamma^\alpha{}\!_{\mu\nu}\dfrac{\text d x^\mu}{\text d s} \dfrac{\text d x^\nu}
{\text d s}=- g^{\alpha\beta} v_i f^i{}\!_{\beta\nu} \dfrac{\text d x^\nu}{\text d s} \quad .
$$
Finally, we define
$f_{\mu\nu}=v_i f^i{}\!_{\mu\nu},$
and see that the above equation becomes
$$
\frac{\text d^2 x^\alpha}{\text d s^2}+ \Gamma^\alpha{}\!_{\mu\nu} \, \frac{\text d x^\mu}{\text d s} \, \frac{\text d
x^\nu}{\text d s} = -f^\alpha{}\!_\nu \frac{\text d x^\nu}{\text d s} \quad , 
$$
which is our Eq.~(\ref{E:C}).\newline

\paragraph{\bf{Equations (\ref{E:D}) and (\ref{E:BBBB})}}

The transformation law for a vector density of weight $+1$ is our Eq.~(\ref{E:D})
$$
V^{\widetilde\alpha}
 =\tfrac{\partial x}{\partial {\widetilde x}} \, x^{\widetilde\alpha}{}\!_\mu V^\mu \quad ,  
$$
Upon differentiating Eq.~(\ref{E:D}) with
respect to $x^{\widetilde\alpha}$, we have
\begin{equation}\label{E:E}
V^{\widetilde\alpha}{}\!,_{\widetilde\alpha}
=\tfrac{\partial x}{\partial {\widetilde x}} V^\alpha{}\!,{}\!_\alpha   +V^\mu \, \left(\tfrac{\partial x}{\partial {\widetilde x}} \, x^{\widetilde\alpha}\!{}_\mu\right)\!,_{\widetilde\alpha}  .
\end{equation}
Now,
$$
\aligned
\left({\tfrac{\partial x}{\partial {\widetilde x}}} \, x^{\widetilde\alpha}\!_\mu\right)\!,\!_{\widetilde\alpha}
&= \left({\tfrac{\partial x}{\partial {\widetilde x}}}\right)\!,_{\widetilde\alpha} x^{\widetilde\alpha}{}\!_\mu   + {\tfrac{\partial x}{\partial {\widetilde x}}} x^{\widetilde\alpha}\!_\mu,_{\widetilde\alpha}
= ({\tfrac{\partial x}{\partial {\widetilde x}}}), _\mu
+ \tfrac{\partial x}{\partial {\widetilde x}} x^{\widetilde\alpha}\!_\mu,_\nu x^\nu{}\!_{\widetilde\alpha} \\
&=\tfrac{\partial x}{\partial {\widetilde x}} x^\nu{}\!_{\widetilde\alpha}(x^{\widetilde\alpha}\!_\mu,_\nu  - \, x^{\widetilde\alpha}\!_\nu,_\mu),
\endaligned
$$
where we have used the well-known formula
$
({\tfrac{\partial x}{\partial {\widetilde x}}}), _\mu=\tfrac{\partial x}{\partial {\widetilde x}} \, x^{\widetilde\alpha}\!_\nu  x^\nu\!_{\widetilde\alpha} , _\mu
$
for the derivative of a determinant.
Upon substituting the expression
$\left({\tfrac{\partial x}{\partial {\widetilde x}}} \, x^{\widetilde\alpha}\!_\mu\right)\!,\!_{\widetilde\alpha}
=\tfrac{\partial x}{\partial {\widetilde x}} x^\nu{}\!_{\widetilde\alpha}(x^{\widetilde\alpha}\!_\mu,_\nu  - \, x^{\widetilde\alpha}\!_\nu,_\mu)$
into Eq.~(\ref{E:E}), we obtain our Eq.~(\ref{E:BBBB}),
$$
V^{\widetilde\alpha}{}\!,_{\widetilde\alpha}
=\tfrac{\partial x}{\partial {\widetilde x}} V^\alpha{}\!,{}\!_\alpha +V^\mu \, \tfrac{\partial x}{\partial {\widetilde x}} x^\nu{}\!_{\widetilde\alpha}(x^{\widetilde\alpha}\!_\mu,_\nu-\, x^{\widetilde\alpha}\!_\nu,_\mu)\, .
$$
\paragraph{\bf{Equations (\ref{E:F}) and (\ref{E:G})}} Upon differentiating Eq.~(\ref{E:F}),
$
x^{\widehat\alpha}\!_\mu = x^{\widehat\alpha}\!_{\widetilde\rho} \, x^{\widetilde\rho}\!_\mu
$
with respect to $x^\nu ,$ we get
$$
x^{\widehat\alpha}\!_\mu ,_\nu = x^{\widehat\alpha}\!_{\widetilde\rho} ,_\nu \, x^{\widetilde\rho}\!_\mu + x^{\widehat\alpha}\!_{\widetilde\rho} \, x^{\widetilde\rho}\!_\mu , _\nu
                = x^{\widehat\alpha}\!_{\widetilde\rho} ,_{\widetilde\sigma} \, x^{\widetilde\rho}\!_\mu x^{\widetilde\sigma}\!_\nu + x^{\widehat\alpha}\!_{\widetilde\rho} \, x^{\widetilde\rho}\!_\mu , _\nu
$$
By subtracting the corresponding expression for $x^{\widehat\alpha}\!_\nu ,_\mu$, and multiplying by
$x^\nu{}\!_{{\widehat\alpha}}$, we get

$$
x^\nu{}\!_{{\widehat\alpha}}(x^{\widehat\alpha}\!_\mu ,_\nu-x^{\widehat\alpha}\!_\nu ,_\mu) = x^{{\widetilde\sigma}}{}\!_{{\widehat\alpha}} (x^{\widehat\alpha}\!_{\widetilde\rho} ,_{\widetilde\sigma} -x^{\widehat\alpha}\!_{\widetilde\sigma} ,_{\widetilde\rho}) x^{\widetilde\rho}\!_\mu
                                 + x^\nu{}\!_{\widetilde\rho} (x^{\widetilde\rho}\!_\mu , _\nu - x^{\widetilde\rho}\!_\nu , _\mu )\; ,
$$
which is Eq.~(\ref{E:G}).

\paragraph{\bf{Validity of Equation (\ref{E:H})}} If we take the partial derivative of the equation
$
x^{\widetilde\alpha}\!_\mu = \delta^\alpha_\mu + \delta^\alpha_0 \delta^2_\mu x^1
$
with respect to $x^\nu$, we get
$
x^{\widetilde\alpha}\!_\mu , _\nu = \delta^\alpha_0 \delta^2_\mu \delta^1_\nu \, .
$
Subtraction of the corresponding expression with $\mu$ and $\nu$ interchanged gives
\begin{equation}\label{E:CCCC}
x^{\widetilde\alpha}\!_\mu , _\nu - x^{\widetilde\alpha}\!_\nu , _\mu  = \delta^\alpha_0 (\delta^2_\mu \delta^1_\nu - \delta^2_\nu \delta^1_\mu)\,.
\end{equation}
If we choose ${\widetilde\alpha} = 0,$ \, $\mu =2,$ \, and $\nu=1$ we get
$$
x^{\widetilde 0}\!_2 , _1 - x^{\widetilde 0}\!_1 , _2  = \delta^0_0 (\delta^2_2 \delta^1_1 - \delta^2_1 \delta^1_2) = 1 \quad ,
$$
which shows that the transformation coefficients in Eq.~(\ref{E:H}) {\sl do not} satisfy Eq.~(\ref{E:A}). However, upon multiplying
Eq.~(\ref{E:CCCC}) by $x^\nu\!_{\widetilde\alpha} = \delta^\nu_\alpha - \delta^\nu_0 \delta^2_\alpha x^1\, ,$ we get
$$
x^\nu\!_{\widetilde\alpha}  (x^{\widetilde\alpha}\!_\mu , _\nu - x^{\widetilde\alpha}\!_\nu , _\mu)  =  \delta^\nu_\alpha\delta^\alpha_0 (\delta^2_\mu \delta^1_\nu - \delta^2_\nu \delta^1_\mu)
                                                     - \delta^\nu_0 \delta^2_\alpha x^1\delta^\alpha_0 (\delta^2_\mu \delta^1_\nu - \delta^2_\nu \delta^1_\mu)=0 \,
$$
which shows that the transformation coefficients in Eq.~(\ref{E:H}) {\sl    do} satisfy Eq.~(\ref{E:B}).\newline

\paragraph{\bf{Equations (\ref{E:K}) and (\ref{E:L})}} From
$
h^i{}\!_ \mu = h^i{}\!_{{\widetilde\alpha}}\, x^{{\widetilde\alpha}} {}\!_\mu
$
we see that
$$
h^i{}\!_ \mu , _\nu = h^i{}\!_{{\widetilde\alpha}}, _\nu x^{{\widetilde\alpha}} {}\!_\mu + h^i{}\!_{{\widetilde\alpha}} x^{{\widetilde\alpha}} {}\!_\mu , _\nu
                  = h^i{}\!_{{\widetilde\alpha}}, _{{\widetilde\beta}} x^{{\widetilde\alpha}} {}\!_\mu  x^{{\widetilde\beta}} {}\!_\nu  + h^i{}\!_{{\widetilde\alpha}} x^{{\widetilde\alpha}} {}\!_\mu , _\nu \, .
$$
If we subtract the corresponding expression for $h^i{}\!_ \nu , _\mu$, and multiply by
$
h_i{}\!^\nu = h_i{}\!^{{\widetilde\sigma}} x^\nu{}\!_{{\widetilde\sigma}} \, ,
$, we get
$$
h_i{}\!^\nu(h^i{}\!_ \mu , _\nu - h^i{}\!_ \nu , _\mu)
  = h_i{}\!^{{\widetilde\sigma}} x^\nu{}\!_{{\widetilde\sigma}}(h^i{}\!_{{\widetilde\alpha}}, _{{\widetilde\beta}}  - h^i{}\!_{{\widetilde\beta}}, _{{\widetilde\alpha}}) x^{{\widetilde\alpha}} {}\!_\mu  x^{{\widetilde\beta}} {}\!_\nu
    + h_i{}\!^\nu h^i{}\!_{{\widetilde\alpha}} (x^{{\widetilde\alpha}} {}\!_\mu , _\nu  -  x^{{\widetilde\alpha}} {}\!_\nu , _\mu) \; .
$$
Thus, from the chain rule
$$
h_i{}\!^\nu(h^i{}\!_ \mu , _\nu - h^i{}\!_ \nu , _\mu)
  = h_i{}\!^{{\widetilde\beta}}(h^i{}\!_{{\widetilde\alpha}}, _{{\widetilde\beta}}  - h^i{}\!_{{\widetilde\beta}}, _{{\widetilde\alpha}}) x^{{\widetilde\alpha}} {}\!_\mu
    +  x^\nu{}\!_{{\widetilde\alpha}} (x^{{\widetilde\alpha}} {}\!_\mu , _\nu  -  x^{{\widetilde\alpha}} {}\!_\nu , _\mu) \; .
$$
This may be written
$$
C_\mu=C_{{\widetilde\alpha}} x^{{\widetilde\alpha}}{}\!_\mu
+ x^\nu{}\!_{{\widetilde\alpha}}\left(x^{{\widetilde\alpha}}{}\!_\mu ,_\nu - x^{{\widetilde\alpha}}{}\!_\nu ,_\mu\right) \quad  
$$
which is our Eq.~(\ref{E:L}).

\paragraph{\bf{Derivation of Equation (\ref{E:i})}} The antisymmetry of $\gamma_{\mu\nu i}$ in $\mu$ and $\nu$  may be used to obtain an expression for
$
\gamma_{\mu\nu i}
$
in terms of
$
f_{i\mu\nu}
$.
We have
$
f_{i\mu\nu}=h_{i\nu},_\mu - h_{i\mu},_\nu =  h_{i\nu ;\mu} - h_{i\mu ;\nu}
$,
so that
$$
f_{i\mu\nu} = \gamma_{i \nu\mu} - \gamma_{i \mu\nu} \quad .
$$
If we subtract the corresponding expressions for $f_{\mu\nu i}$ and $f_{\nu i\mu}$, we have
$$
\aligned
f_{i\mu\nu}-f_{\mu\nu i}-f_{\nu i\mu}=&\gamma_{i \nu\mu}-\gamma_{i \mu\nu}-\gamma_{\mu i\nu}+\gamma_{\mu \nu i}-\gamma_{\nu \mu i}+\gamma_{\nu i \mu}\\
                              =-&\gamma_{\nu i\mu}-\gamma_{i \mu\nu}+\gamma_{i\mu\nu}+\gamma_{\mu \nu i}+\gamma_{\mu \nu i}+\gamma_{\nu i \mu}=2\gamma_{\mu\nu i}
\endaligned
$$
so
$$
\gamma_{\mu\nu i} = \tfrac{1}{2} \left(f_{i\mu\nu} - f_{\mu\nu i} - f_{\nu i\mu}\right) 
\;
$$
which is our Eq.~(\ref{E:i}).\newline

\paragraph{\bf{Equations (\ref{E:nnn}) and (\ref{E:g})}}
We note that $\sqrt{-g}$ equals $h$, the determinant of $h^i{}\!_\mu\, ,$
so we may write Eq.~(\ref{E:nnn}) as
\begin{equation}\label{E:ooo}
0=\int h\left(2\, C^j C_\nu - \,C^\alpha C_\alpha h^j{}\!_\nu\right) \delta h_j{}\!^\nu  \text d^4\! x +\int 2h C^\mu \delta C_\mu  \text d^4\! x  \quad .
\end{equation}
We now need an expression for $\delta C_\mu$ to use in the last integral of Eq.~(\ref{E:ooo}).\newline
From
$
C_\mu=h_j{}^\nu {} \left(h^j{}\!_\mu ,_\nu-h^j{}\!_\nu ,_\mu\right) \, ,
$
we have
$$
\aligned
\delta C_\mu&=\left(h^j{}\!_\mu ,_\nu-h^j{}\!_\nu ,_\mu\right) \delta h_j{}^\nu + h_j{}^\nu \left[  (\delta h^j{}\!_\mu) ,_\nu-(\delta h^j{}\!_\nu) ,_\mu\right] \\
             &=\left(\gamma^j{}\!_{\mu\nu}-\gamma^j{}\!_{\nu\mu}\right) \delta h_j{}^\nu + h_j{}^\nu (\delta h^j{}\!_\mu)_{;\nu}- h_j{}^\nu(\delta h^j{}\!_\nu)_{;\mu}\; .
\endaligned
$$
Upon multiplying by
$
2hC^\mu
$,
we get an expression which we manipulate into the form
$$
\aligned
2hC^\mu \delta C_\mu=&h\left(2C^\mu \gamma^j{}\!_{\mu\nu}-2C^\mu \gamma^j{}\!_{\nu\mu}\right) \delta h_j{}^\nu \\
   &+h\left(2C^\alpha{}\!_{;\alpha} h_i{}^\beta + 2C^\mu \gamma_i{}^\beta{}\!_\mu  -2C^\beta{}\!_{;\mu} h_i{}^\mu -2C^\beta \gamma_i{}^\mu{}\!_\mu\right)
    \delta h^i{}\!_\beta \\
               &+\left(2hC^\mu h_j{}^\nu \delta h^j{}\!_\mu
               -2hC^\nu h_j{}^\mu \delta h^j{}\!_\mu\right)\! ,_\nu \quad .
\endaligned
$$
We now need an expression for $\delta h^i{}\!_\beta$ in terms of $\delta h_j{}\!^\nu$. From $h^j{}\!_\beta \, h_j{}\!^\nu=\delta^\nu_\beta$ we get
$h_j{}\!^\nu \, \delta h^j{}\!_\beta+h^j{}\!_\beta \,\delta h_j{}\!^\nu = 0 \, ,$ so
$h^i{}\!_\nu \, h_j{}\!^\nu \, \delta h^j{}\!_\beta+ h^i{}\!_\nu \,  h^j{}\!_\beta \,\delta h_j{}\!^\nu = 0 \, .$ Thus, we find that
$\delta h^i{}\!_\beta = -h^i{}\!_\nu \, h^j{}\!_\beta \,\delta h_j{}\!^\nu \, .$
Substituting this expression for $\delta h^i{}\!_\beta$ in the above expression for $2hC^\mu \delta C_\mu$ gives
$$
\aligned
2hC^\mu \delta C_\mu=&h\left(2C^\mu \gamma^j{}\!_{\mu\nu}-2C^\mu \gamma^j{}\!_{\nu\mu}\right) \delta h_j{}^\nu \\
               &-h\left(2C^\alpha{}\!_{;\alpha} h_i{}^\beta h^i{}\!_\nu \, h^j{}\!_\beta
                  + 2C^\mu \gamma_i{}^\beta{}\!_\mu h^i{}\!_\nu \, h^j{}\!_\beta\right) \delta h_j{}\!^\nu \\
               &+h\left(2C^\beta{}\!_{;\mu} h_i{}^\mu h^i{}\!_\nu \, h^j{}\!_\beta
                  + 2C^\beta \gamma_i{}^\mu{}\!_\mu h^i{}\!_\nu \, h^j{}\!_\beta\right) \delta h_j{}\!^\nu \\
               &+\left(2hC^\mu h_j{}^\nu \delta h^j{}\!_\mu
               -2hC^\nu h_j{}^\mu \delta h^j{}\!_\mu\right)\! ,_\nu \\
                =&h\left(2C^\mu \gamma^j{}\!_{\mu\nu}-2C^\mu \gamma^j{}\!_{\nu\mu}\right) \delta h_j{}^\nu \\
               &-h\left(2C^\alpha{}\!_{;\alpha} \, h^j{}\!_\nu
                  - 2C^\mu  \, \gamma^j{}\!_{\nu\mu}\right) \delta h_j{}\!^\nu \\
               &+h\left(2C^\beta{}\!_{;\nu}  \, h^j{}\!_\beta
                  - 2C^j  \, \gamma^\mu{}\!_{\nu\mu}\right) \delta h_j{}\!^\nu \\
               &+\left(2hC^\mu h_j{}^\nu \delta h^j{}\!_\mu
               -2hC^\nu h_j{}^\mu \delta h^j{}\!_\mu\right)\! ,_\nu
\endaligned
$$
We use $\gamma^\mu{}\!_{\nu\mu}=C_\nu$, and cancellations to obtain
$$
\aligned
2hC^\mu \delta C_\mu=
&h\left(2C^\beta{}\!_{;\nu}\, h^j{}\!_\beta-2C^j\, C_\nu +2C^\mu \gamma^j{}\!_{\mu\nu}-2C^\alpha{}\!_{;\alpha} \, h^j{}\!_\nu   \right) \delta h_j{}^\nu \\               &+\left(2hC^\mu h_j{}^\nu \delta h^j{}\!_\mu
               -2hC^\nu h_j{}^\mu \delta h^j{}\!_\mu\right)\! ,_\nu \, .
\endaligned
$$
Thus, we have
$$
\aligned
\int 2hC^\mu \delta C_\mu  \text d^4x =
&\int h\left(2C^\beta{}\!_{;\nu}\, h^j{}\!_\beta-2C^j\,C_\nu +2C^\mu \gamma^j{}\!_{\mu\nu}-2C^\alpha{}\!_{;\alpha}\,h^j{}\!_\nu   \right) \delta h_j{}^\nu  \text d^4x \\
&+\int\left(2hC^\mu h_j{}^\nu \delta h^j{}\!_\mu
               -2hC^\nu h_j{}^\mu \delta h^j{}\!_\mu\right)\! ,_\nu \,  \text d^4x \, ,
\endaligned
$$
and using this expression for $\int 2hC^\mu \delta C_\mu d^4x$ in Eq.~(\ref{E:ooo}) gives
$$
\aligned
0&=\int h\left(2C^\beta{}\!_{;\nu}\, h^j{}\!_\beta +2C^\alpha \gamma^j{}\!_{\alpha\nu}-2C^\alpha{}\!_{;\alpha}\,h^j{}\!_\nu -C^\alpha C_\alpha h^j{}\!_\nu \right) \delta h_j{}^\nu  \text d^4x \\      &+\int\left[ 2h(C^\mu h_j{}^\nu -C^\nu h_j{}^\mu)\delta h^j{}\!_\mu \right]\! ,_\nu \,  \text d^4x \quad .
\endaligned
$$
By using Gauss's theorem, we may write the second integral above
as an integral over the boundary of the region of integration. We discard this boundary integral by demanding that
$\delta h^j{}\!_\mu$ shall vanish on the boundary. Thus, we have
$$
0=\int h\left(2C^\beta{}\!_{;\nu}\, h^j{}\!_\beta +2C^\alpha \gamma^j{}\!_{\alpha\nu}-2C^\alpha{}\!_{;\alpha}\,h^j{}\!_\nu -C^\alpha C_\alpha h^j{}\!_\nu \right)
\delta h_j{}^\nu  \text d^4x  \quad ,
$$
and by allowing the variation $\delta h_j{}^\nu$ to be arbitrary in the interior of the region of integration, we obtain
the Euler-Lagrange equations
$$
C^\beta{}\!_{;\nu}\, h^j{}\!_\beta +C^\alpha \gamma^j{}\!_{\alpha\nu}-C^\alpha{}\!_{;\alpha}\,h^j{}\!_\nu - \tfrac{1}{2} C^\alpha C_\alpha h^j{}\!_\nu = 0
$$
which are our field equations, but not in the form that we desire. If we multiply by $h_j{}\!^\sigma$$g_{\sigma\mu}$, we get
$$
C_{\mu ;\nu}-C_\alpha \gamma^\alpha{}\!_{\mu\nu}- g_{\mu\nu} C^\alpha{}\!_{;\alpha} - \tfrac{1}{2} g_{\mu\nu}  C^\alpha C_\alpha = 0 \quad .
$$
This is our Eq.~(\ref{E:g}), i.e., our field equations as given in Sec.~5.1.
\paragraph{\bf{Noether Conserved Currents: The Antisymmetric Part of the Field Equations}} We now apply Noether's
theorem to the integral which was discarded after using Gauss's theorem. The integrand of this integral is
$$
2h(C^\mu h_j{}^\nu -C^\nu h_j{}^\mu)\delta h^j{}\!_\mu \, .
$$
A variation
$
\delta h^j{}\!_\mu=g^{jn}\, \varepsilon_{nk}\, h^k{}\!_\mu \, ,
$
where the $\varepsilon_{nk}$ are infinitesimal antisymmetric constants, leaves the metric unchanged. From this, and the straightforward use of Noether's theorem, we find that the six currents
$$
{\frak J}_{ij}{}\!^\nu = C_\alpha(h_i{}\!^\alpha \, h_j{}\!^\nu - h_j{}\!^\alpha \, h_i{}\!^\nu)
$$
satisfy the conservation laws
$
{\frak J}_{ij}{}\!^\nu{}\!_{:\nu}=0 \, .
$
We note that
$
{\frak J}_{ij}{}\!^\nu
$
is a vector under conservative coordinate transformations from $x^\alpha$ to $x^{\widetilde\alpha}$, and that these ``proper''
conservation laws may be expressed in the form
$
(\sqrt{-g} {\frak J}_{ij}{}\!^\nu),_\nu =0.
$
They are ``weak'' conservation laws in the sense that
$
{\frak J}_{ij}{}\!^\nu
$
is conserved if the field equations are satisfied. Indeed, the antisymmetric part of Eq.(\ref{E:g}) just states that
$
{\frak J}_{ij}{}\!^\nu{}\!_{:\nu}=0 \, .
$
\newline

\paragraph{\bf {Derivation of Eq.~(\ref{E:AAAA})}}

Our identity for the Einstein tensor, Eq.~(\ref{E:T}), is
$$
\begin{aligned}
G_{\mu\nu} =& \,C_{\mu;\nu}-C_\alpha \gamma^\alpha{}\!_{\mu\nu}
     -g_{\mu\nu} C^\alpha{}\!_{;\alpha}
   -\tfrac{1}{2} g_{\mu\nu} C^\alpha C_\alpha \\
          & + \gamma_\mu{}\!^\alpha{}\!_{\nu;\alpha} + \gamma^\alpha{}\!_{\sigma\nu} \gamma^\sigma{}\!_{\mu\alpha}
              + \tfrac{1}{2} g_{\mu\nu} \gamma^{\alpha\sigma\beta} \gamma_{\beta\sigma\alpha} \quad , 
\end{aligned}
$$
and we subtract the corresponding expression for $G_{\nu\mu}$ to obtain
\begin{equation}\label{E:sss}
\aligned
G_{\mu\nu}-G_{\nu\mu} =& (C_{\mu;\nu}-C_\alpha \gamma^\alpha{}\!_{\mu\nu})
                     -(C_{\nu;\mu}-C_\alpha \gamma^\alpha{}\!_{\nu\mu}) \\
         &+ (\gamma_\mu{}\!^\alpha{}\!_{\nu;\alpha}+ \gamma^\alpha{}\!_{\sigma\nu} \gamma^\sigma{}\!_{\mu\alpha})
                  - (\gamma_\nu{}\!^\alpha{}\!_{\mu;\alpha}
                + \gamma^\alpha{}\!_{\sigma\mu} \gamma^\sigma{}\!_{\nu\alpha}) \quad .
\endaligned
\end{equation}
It is obvious that $G_{\mu\nu}-G_{\nu\mu}=R_{\mu\nu}-R_{\nu\mu},$ so we seek an identity for $R_{\mu\nu}-R_{\nu\mu}$ which is valid
if the tetrad is path-dependent.

The covariant derivative of $h^i{}\!_\beta$ with respect to $x^\mu$ is given by
\begin{equation}\label{E:oo}
h^i{}\!_{\beta;\mu}= h^i{}\!_\beta,_\mu - h^i{}\!_\sigma \Gamma^\sigma{}\!_{\beta\mu}
\end{equation}
where
$
\Gamma^\alpha{}\!_{\mu\nu}= \tfrac{1}{2}g^{\alpha\beta}\left(g_{\beta\mu},_\nu + g_{\beta\nu},_\mu - g_{\mu\nu},_\beta\right)
$
is the usual Christoffel symbol. The covariant derivative of Eq.~(\ref{E:oo}) with respect to $x^\nu$ gives
\begin{equation}\label{E:pp}
\aligned
h^i{}\!_{\beta;\mu;\nu}=&h^i{}\!_\beta ,_\mu\! ,_\nu -h^i{}\!_\sigma,_\nu \Gamma^\sigma{}\!_{\beta\mu}-h^i{}\!_\sigma \Gamma^\sigma{}\!_{\beta\mu},_\nu
                                          -h^i{}\!_\sigma,_\mu \Gamma^\sigma{}\!_{\beta\nu}\\
                                 &+h^i{}\!_\sigma \Gamma^\sigma{}\!_{\rho\mu} \Gamma^\rho{}\!_{\beta\nu}
                                 -h^i{}\!_\beta,_\sigma \Gamma^\sigma{}\!_{\mu\nu}+h^i{}\!_\sigma \Gamma^\sigma{}\!_{\beta\rho} \Gamma^\rho{}\!_{\mu\nu}\quad .
\endaligned
\end{equation}
If we subtract from Eq.~(\ref{E:pp}) the corresponding expression with $\mu$ and $\nu$ interchanged, we get
$$
h^i{}\!_{\beta;\mu;\nu}-h^i{}\!_{\beta;\nu;\mu}= \left[\partial_\nu , \partial_\mu\right] h^i{}\!_\beta
                       +h^i{}\!_\sigma \left(\Gamma^\sigma{}\!_{\beta\nu},_\mu-\Gamma^\sigma{}\!_{\beta\mu},_\nu
                       +\Gamma^\sigma{}\!_{\rho\mu} \Gamma^\rho{}\!_{\beta\nu}-\Gamma^\sigma{}\!_{\rho\nu} \Gamma^\rho{}\!_{\beta\mu}\right)
                                 \quad ,
$$
and if we multiply this by $h_i{}\!^\alpha$, we obtain
\begin{equation}\label{E:qq}
R^\alpha{}\!_{\beta\mu\nu}= h_i{}\!^\alpha \left[\partial_\nu , \partial_\mu\right] h^i{}\!_\beta +{\Bbb R}^\alpha{}\!_{\beta\mu\nu}
\end{equation}
where
\begin{equation}\label{E:rr}
{\Bbb R}^\alpha{}\!_{\beta\mu\nu}=\Gamma^\alpha{}\!_{\beta\nu},_\mu-\Gamma^\alpha{}\!_{\beta\mu},_\nu
                       +\Gamma^\alpha{}\!_{\rho\mu} \Gamma^\rho{}\!_{\beta\nu}-\Gamma^\alpha{}\!_{\rho\nu} \Gamma^\rho{}\!_{\beta\mu}
\end{equation}
is the orthodox expression for the Riemann tensor in terms of the metric, rather than in terms of the tetrad.
We see from Eq.~(\ref{E:qq}) that
\begin{equation}\label{E:ss}
R_{\mu\nu}= h_i{}\!^\alpha \left[\partial_\nu , \partial_\alpha\right] h^i{}\!_\mu +{\Bbb R}_{\mu\nu}
\end{equation}
where
$
{\Bbb R}_{\mu\nu}={\Bbb R}^\alpha{}\!_{\mu\alpha\nu}\, .
$
Thus
\begin{equation}\label{E:tt}
R_{\mu\nu}-R_{\nu\mu} = h_i{}\!^\alpha \left(\left[\partial_\nu , \partial_\alpha\right] h^i{}\!_\mu + \left[\partial_\alpha , \partial_\mu\right] h^i{}\!_\nu\right)
                   +{\Bbb R}_{\mu\nu}-{\Bbb R}_{\nu\mu} \quad .
\end{equation}

We now compute the identity for
$
{\Bbb R}_{\mu\nu}-{\Bbb R}_{\nu\mu} \, .
$
We see from Eq.~(\ref{E:rr}) that
$$
{\Bbb R}_{\alpha\beta\mu\nu}= g_{\alpha\sigma} \Gamma^\sigma{}\!_{\beta\nu},_\mu-g_{\alpha\sigma}\Gamma^\sigma{}\!_{\beta\mu},_\nu
                       +[\alpha,\rho\mu]\Gamma^\rho{}\!_{\beta\nu}-[\alpha,\rho\nu] \Gamma^\rho{}\!_{\beta\mu}
$$
where
$
[\alpha,\mu\nu]= \tfrac{1}{2}\left(g_{\alpha\mu},_\nu + g_{\alpha\nu},_\mu - g_{\mu\nu},_\alpha\right).
$
Thus, we have
\begin{equation}\label{E:uu}
\aligned
{\Bbb R}_{\alpha\beta\mu\nu}= &\left(g_{\alpha\sigma} \Gamma^\sigma{}\!_{\beta\nu}\right)\!,_\mu -g_{\alpha\sigma},_\mu \Gamma^\sigma{}\!_{\beta\nu}
                    -\left(g_{\alpha\sigma} \Gamma^\sigma{}\!_{\beta\mu}\right)\!,_\nu +g_{\alpha\sigma},_\nu \Gamma^\sigma{}\!_{\beta\mu}\\
                     &+[\alpha,\rho\mu]\Gamma^\rho{}\!_{\beta\nu}-[\alpha,\rho\nu] \Gamma^\rho{}\!_{\beta\mu}
                     +[\alpha,\sigma\mu] \Gamma^\sigma{}\!_{\beta\nu} - [\alpha,\sigma\nu] \Gamma^\sigma{}\!_{\beta\mu} \\
                   = &[\alpha,\beta\nu]\!,_\mu - [\alpha,\beta\mu]\!,_\nu
                      +([\alpha,\sigma\mu]-g_{\alpha\sigma},_\mu)\Gamma^\sigma{}\!_{\beta\nu} \\
                      &-([\alpha,\sigma\nu]-g_{\alpha\sigma},_\nu)\Gamma^\sigma{}\!_{\beta\mu} .
\endaligned
\end{equation}
Now,
\begin{equation}\label{E:vv}
\left[\alpha,\sigma\mu\right] -g_{\alpha\sigma}\!,_\mu = \tfrac{1}{2}\left(g_{\alpha\sigma},_\mu + g_{\alpha\mu},_\sigma - g_{\sigma\mu},_\alpha\right)- g_{\alpha\sigma}\!,_\mu =-\left[\sigma,\alpha\mu\right] \, .
\end{equation}
From Eq.~(\ref{E:vv}), we have
$
\left[\alpha,\sigma\mu\right] -g_{\alpha\sigma}\!,_\mu=-\left[\sigma,\alpha\mu\right] \, ,
$
and similarly\newline
$
\left[\alpha,\sigma\nu\right] -g_{\alpha\sigma}\!,_\nu=-\left[\sigma,\alpha\nu\right] \, .
$
Upon using these expressions in Eq.~(\ref{E:uu}), we have
$$
{\Bbb R}_{\alpha\beta\mu\nu}=[\alpha,\beta\nu]\!,_\mu - [\alpha,\beta\mu]\!,_\nu - \left[\sigma,\alpha\mu\right]\Gamma^\sigma{}\!_{\beta\nu}+\left[\sigma,\alpha\nu\right]\Gamma^\sigma{}\!_{\beta\mu}
$$
so that
\begin{equation}\label{E:ww}
{\Bbb R}_{\alpha\mu\beta\nu}=[\alpha,\mu\nu]\!,_\beta - [\alpha,\beta\mu]\!,_\nu - \left[\sigma,\alpha\beta\right]\Gamma^\sigma{}\!_{\mu\nu}+\left[\sigma,\alpha\nu\right]\Gamma^\sigma{}\!_{\beta\mu} \quad .
\end{equation}
Upon multiplying Eq.~(\ref{E:ww}) by $g^{\alpha\beta}$, we obtain
\begin{equation}\label{E:xx}
{\Bbb R}_{\mu\nu}=g^{\alpha\beta}[\alpha,\mu\nu]\!,_\beta - g^{\alpha\beta}[\alpha,\beta\mu]\!,_\nu
                - g^{\alpha\beta}\left[\sigma,\alpha\beta\right]\Gamma^\sigma{}\!_{\mu\nu}+g^{\alpha\beta}\left[\sigma,\alpha\nu\right]\Gamma^\sigma{}\!_{\beta\mu} \quad .
\end{equation}
We see from Eq.~(\ref{E:xx}) that
$$
{\Bbb R}_{\nu\mu}=g^{\alpha\beta}[\alpha,\nu\mu]\!,_\beta - g^{\alpha\beta}[\alpha,\beta\nu]\!,_\mu
                - g^{\alpha\beta}\left[\sigma,\alpha\beta\right]\Gamma^\sigma{}\!_{\nu\mu}+g^{\alpha\beta}\left[\sigma,\alpha\mu\right]\Gamma^\sigma{}\!_{\beta\nu} \quad .
$$
We use the facts that $[\alpha,\nu\mu]=[\alpha,\mu\nu]$ and that $\Gamma^\sigma{}\!_{\nu\mu}=\Gamma^\sigma{}\!_{\mu\nu}$. Also, in the last term above,
change $\alpha$ to $\beta$ and $\beta$ to $\alpha$. Thus, we obtain
$$
{\Bbb R}_{\nu\mu}=g^{\alpha\beta}[\alpha,\mu\nu]\!,_\beta - g^{\alpha\beta}[\alpha,\beta\nu]\!,_\mu
                - g^{\alpha\beta}\left[\sigma,\alpha\beta\right]\Gamma^\sigma{}\!_{\mu\nu}+g^{\alpha\beta}\left[\sigma,\beta\mu\right]\Gamma^\sigma{}\!_{\alpha\nu} \quad ,
$$
Next, we use $\Gamma^\sigma{}\!_{\alpha\nu}=g^{\sigma\rho}[\rho,\alpha\nu]$ to obtain
\begin{equation}\label{E:yy}
{\Bbb R}_{\nu\mu}=g^{\alpha\beta}[\alpha,\mu\nu]\!,_\beta - g^{\alpha\beta}[\alpha,\beta\nu]\!,_\mu
                - g^{\alpha\beta}\left[\sigma,\alpha\beta\right]\Gamma^\sigma{}\!_{\mu\nu}+g^{\alpha\beta}\left[\sigma,\alpha\nu\right]\Gamma^\sigma{}\!_{\beta\mu} \quad .
\end{equation}
Upon subtracting Eq.~(\ref{E:yy}) from Eq.~(\ref{E:xx}), we get
$$
\aligned
{\Bbb R}_{\mu\nu}-{\Bbb R}_{\nu\mu}&=g^{\alpha\beta}([\alpha,\beta\nu]\!,_\mu-[\alpha,\beta\mu]\!,_\nu)\\
                  &=\tfrac{1}{2}g^{\alpha\beta}g_{ij}h^j{}\!_\beta[\partial_\mu ,\partial_\nu]h^i{}\!_\alpha
                 +\tfrac{1}{2}g^{\alpha\beta}g_{ij}h^i{}\!_\alpha[\partial_\mu ,\partial_\nu]h^j{}\!_\beta
  \endaligned
$$
Thus, with some obvious raising and lowering of indices, and a renaming of summed indices we obtain
\begin{equation}\label{E:zz}
{\Bbb R}_{\mu\nu}-{\Bbb R}_{\nu\mu}= h_i{}\!^\alpha[\partial_\mu ,\partial_\nu]h^i{}\!_\alpha \quad .
\end{equation}
We now use Eq.~(\ref{E:zz}) in Eq.~(\ref{E:tt}) to obtain
\begin{equation}\label{E:rrr}
R_{\mu\nu}-R_{\nu\mu}=G_{\mu\nu}-G_{\nu\mu} = h_i{}\!^\alpha \left(\left[\partial_\nu , \partial_\alpha\right] h^i{}\!_\mu + \left[\partial_\alpha , \partial_\mu\right] h^i{}\!_\nu
                     + [\partial_\mu ,\partial_\nu]h^i{}\!_\alpha \right)
                    \quad ,
\end{equation}
which is our desired identity, Eq.~(\ref{E:AAAA}).

Consider a tetrad which satisfies the field equations and is path-independent in some $x^\alpha$ coordinate system. If this tetrad is transformed by a conservative transformation to an $x^{\widetilde\alpha}$ coordinate system, the transformed tetrad also satisfies the field equations, but is generally path-dependent. Thus, any statement that a tetrad is, or is not, path-independent has no invariant meaning. As stated in Sec.~5.2, we suspect
that any tetrad which satisfies the field equations admits a conservative transformation to a coordinate system in which
it is path-independent, but we have not succeeded in proving this.\newline

\paragraph{\bf {Equations (\ref{E:yyy}) and (\ref{E:EE})}}
The analysis for manipulating the right-side of Eq.~(\ref{E:yyy}) into a form such that its physical significance is transparent is without approximation, but is rather tedious and not entirely straightforward. Therefore, we now give this analysis in some detail:

We see from Eq.~(\ref{E:j}) and Eq.~(\ref{E:k}) that the totally antisymmetric part and the mixed symmetry part of $\gamma_{\mu\alpha\nu}$ are
$
A_{\mu\alpha\nu}=\tfrac{1}{3} (\gamma_{\mu\alpha\nu} + \gamma_{\alpha\nu\mu}+\gamma_{\nu\mu\alpha})
$
and
$
M_{\mu\alpha\nu}=\tfrac{1}{3}(2\gamma_{\mu\alpha\nu}-\gamma_{\alpha\nu\mu} -\gamma_{\nu\mu\alpha}) \, ,
$
respectively, and we recall that
$
M
$
is antisymmetric in its first two indices. Thus, we have
$
\gamma_\mu{}\!^\alpha{}\!_\nu=A_\mu{}\!^\alpha{}\!_\nu+M_\mu{}\!^\alpha{}\!_\nu
$
and
$
\gamma_\nu{}\!^\alpha{}\!_\mu=A_\nu{}\!^\alpha{}\!_\mu+M_\nu{}\!^\alpha{}\!_\mu. \,
$
Since
$
A_\mu{}\!^\alpha{}\!_\nu+A_\nu{}\!^\alpha{}\!_\mu=0 \, ,
$
we see that
$
\gamma_\mu{}\!^\alpha{}\!_\nu+\gamma_\nu{}\!^\alpha{}\!_\mu=M_\mu{}\!^\alpha{}\!_\nu+M_\nu{}\!^\alpha{}\!_\mu \, .
$
By using this in Eq.~(\ref{E:yyy}), we obtain
\begin{equation}\label{E:r}
T_{\mu\nu} =\tfrac{1}{2}(M_\mu{}\!^\alpha{}\!_{\nu;\alpha} +M_\nu{}\!^\alpha{}\!_{\mu;\alpha}
            +\gamma^\alpha{}\!_{\sigma\nu} \gamma^\sigma{}\!_{\mu\alpha}+\gamma^\alpha{}\!_{\sigma\mu} \gamma^\sigma{}\!_{\nu\alpha}
              +  g_{\mu\nu} \gamma^{\alpha\sigma\beta} \gamma_{\beta\sigma\alpha}) \quad .
\end{equation}
Now,
$
M_\mu{}\!^\alpha{}\!_{\nu;\alpha}= \left(M_\mu{}\!^\alpha{}\!_i h^i{}\!_\nu\right)_{;\alpha}
=M_\mu{}\!^\alpha{}\!_{i;\alpha}h^i{}\!_\nu+M_\mu{}\!^\alpha{}\!_i \, h^i{}\!_{\nu;\alpha}=
 M_\mu{}\!^\alpha{}\!_{i;\alpha}h^i{}\!_\nu+M_\mu{}\!^\alpha{}\!_\sigma \, \gamma^\sigma{}\!_{\nu\alpha}
$, so we have
\begin{equation}\label{E:s}
M_\mu{}\!^\alpha{}\!_{\nu;\alpha}=J_{\mu i}\, h^i{}\!_\nu + M_\mu{}\!^\alpha{}\!_\sigma \, \gamma^\sigma{}\!_{\nu\alpha}
\end{equation}
where
$
J_{\mu i}=M_\mu{}\!^\alpha{}\!_{i;\alpha}
$
is the current which is the source of the field
$
M_{\mu\nu i}
$.
Upon using Eq.~(\ref{E:s}) and the corresponding expression for
$
M_\nu{}\!^\alpha{}\!_{\mu;\alpha}
$
in Eq.~(\ref{E:r}) we get
\begin{equation}\label{E:t}
\aligned
T_{\mu\nu} =&\tfrac{1}{2}\left(J_{\mu i}\, h^i{}\!_\nu+J_{\nu i}\, h^i{}\!_\mu + M_\mu{}\!^\alpha{}\!_\sigma \, \gamma^\sigma{}\!_{\nu\alpha}
                               + M_\nu{}\!^\alpha{}\!_\sigma \, \gamma^\sigma{}\!_{\mu\alpha}\right) \\
            &+\tfrac{1}{2}\left(\gamma^\alpha{}\!_{\sigma\nu} \gamma^\sigma{}\!_{\mu\alpha}+\gamma^\alpha{}\!_{\sigma\mu} \gamma^\sigma{}\!_{\nu\alpha}
              +  g_{\mu\nu} \gamma^{\alpha\sigma\beta} \gamma_{\beta\sigma\alpha}\right) \quad .
\endaligned
\end{equation}
We now express each Ricci rotation coefficient in Eq.~(\ref{E:t}) as the sum of its antisymmetric and its mixed-symmetry parts.
Thus, we have
$$
\aligned
T_{\mu\nu} =&\tfrac{1}{2}\left[J_{\mu i}\, h^i{}\!_\nu+J_{\nu i}\, h^i{}\!_\mu
+ M_\mu{}\!^\alpha{}\!_\sigma \, \left( A^\sigma{}\!_{\nu\alpha}+M^\sigma{}\!_{\nu\alpha}\right)
                               + M_\nu{}\!^\alpha{}\!_\sigma \, \left(A^\sigma{}\!_{\mu\alpha}+M^\sigma{}\!_{\mu\alpha}\right)\right] \\
            &+\tfrac{1}{2}\left[\left(A^\alpha{}\!_{\sigma\nu}+M^\alpha{}\!_{\sigma\nu}\right)\left(A^\sigma{}\!_{\mu\alpha}+M^\sigma{}\!_{\mu\alpha}\right)
                                  +\left(A^\alpha{}\!_{\sigma\mu}+M^\alpha{}\!_{\sigma\mu}\right)\left(A^\sigma{}\!_{\nu\alpha}+M^\sigma{}\!_{\nu\alpha}\right)\right] \\
              &+\tfrac{1}{2} \left[g_{\mu\nu}\left(A^{\alpha\sigma\beta}+M^{\alpha\sigma\beta}\right)\left(A_{\beta\sigma\alpha}+M_{\beta\sigma\alpha} \right)\right] \quad .
\endaligned
$$
Upon clearing the inner parentheses, changing the order of terms, and using the total antisymmetry of $A$ and antisymmetry
of $M$ in its first two indices, we obtain after a little factoring
$$
\aligned
T_{\mu\nu} =&\tfrac{1}{2}\left[J_{\mu i}\, h^i{}\!_\nu+J_{\nu i}\, h^i{}\!_\mu
+ M_\mu{}\!^\alpha{}\!_\sigma \left( M^\sigma{}\!_{\nu\alpha}+ \, M_\alpha{}\!^\sigma{}\!_\nu\right)
                                 + M_\nu{}\!^\alpha{}\!_\sigma \left(M^\sigma{}\!_{\mu\alpha}+ M_\alpha{}\!^\sigma{}\!_\mu\right)\right] \\
&+\tfrac{1}{2}\left[A_{\mu\alpha}{}\!^\sigma \left( M_\nu{}\!^\alpha{}\!_\sigma+ M^\alpha{}\!_{\sigma\nu}+ M_{\sigma\nu}{}\!^\alpha\right)
+A_{\nu\alpha}{}\!^\sigma \left( M_\mu{}\!^\alpha{}\!_\sigma+ M^\alpha{}\!_{\sigma\mu}+ M_{\sigma\mu}{}\!^\alpha\right)\right]\\
      &+ A^\alpha{}\!_{\mu\sigma} \, A_{\alpha\nu}{}\!^\sigma
       +\tfrac{1}{2} g_{\mu\nu}\left(2A^{\alpha\sigma\beta}\,M_{\beta\sigma\alpha}
                                             +M^{\alpha\sigma\beta}\,M_{\beta\sigma\alpha}-A^{\alpha\beta\sigma}\,A_{\alpha\beta\sigma}\right) \quad .
\endaligned
$$
It follows from Eq.~(\ref{E:k}) that
\begin{equation}\label{E:n}
M_{\mu\nu i}+M_{i\mu\nu}+M_{\nu i\mu}=0  \; .  
\end{equation}
We easily find from Eqs.~(\ref{E:j}) and (\ref{E:k}) that
\begin{equation}\label{E:o}
M^{\mu\nu\alpha}\, A_{\mu\nu\alpha}=0 \, .
\end{equation}
Now
$$
M^{\alpha\sigma\beta} \, M_{\beta\sigma\alpha}={\tfrac{1}{2}}M^{\alpha\sigma\beta} \, M_{\beta\sigma\alpha}+{\tfrac{1}{2}}M^{\alpha\sigma\beta} \, M_{\beta\sigma\alpha}
={\tfrac{1}{2}}M^{\alpha\sigma\beta} \, M_{\beta\sigma\alpha}+\tfrac{1}{2}M^{\sigma\alpha\beta} \, M_{\beta\alpha\sigma}
$$
and since $\bf M$ is antisymmetric in its first two indices,
$$
M^{\alpha\sigma\beta} \, M_{\beta\sigma\alpha}={\tfrac{1}{2}}M^{\alpha\sigma\beta} \, M_{\beta\sigma\alpha}+{\tfrac{1}{2}}M^{\alpha\sigma\beta} \, M_{\alpha\beta\sigma}
                        ={\tfrac{1}{2}}M^{\alpha\sigma\beta}\left( M_{\beta\sigma\alpha}+M_{\alpha\beta\sigma}\right)\; .
$$
Obviously Eq.~(\ref{E:n}) implies that
$
M_{\beta\sigma\alpha}+M_{\alpha\beta\sigma}=-M_{\sigma\alpha\beta}=M_{\alpha\sigma\beta}
$
so we see that
\begin{equation}\label{E:p}
M^{\alpha\sigma\beta} \, M_{\beta\sigma\alpha}={\frac{1}{2}}M^{\alpha\sigma\beta}M_{\alpha\sigma\beta} \quad . 
\end{equation}
Thus, we have
\begin{equation}\label{E:u}
\aligned
&T_{\mu\nu} =\tfrac{1}{2}\left(J_{\mu i}\, h^i{}\!_\nu+J_{\nu i}\, h^i{}\!_\mu
- M_\mu{}\!^\alpha{}\!_\sigma M_{\nu\alpha}{}\!^\sigma - M_\nu{}\!^\alpha{}\!_\sigma M_{\mu\alpha}{}\!^\sigma \right)
      + A^\alpha{}\!_{\mu\sigma} \, A_{\alpha\nu}{}\!^\sigma \\
       &+\frac{1}{2} g_{\mu\nu}\left(2A^{\alpha\sigma\beta}\,M_{\beta\sigma\alpha}
                                             +M^{\alpha\sigma\beta}\,M_{\beta\sigma\alpha}-A^{\alpha\beta\sigma}\,A_{\alpha\beta\sigma}\right) \quad .
\endaligned
\end{equation}
Therefore, Eq.~(\ref{E:u}) reduces to
\begin{equation}\label{E:v}
\aligned
T_{\mu\nu} =&\tfrac{1}{2}\left(J_{\mu i}\, h^i{}\!_\nu+J_{\nu i}\, h^i{}\!_\mu\right)
 -\left(M_{\mu\alpha}{}\!^\sigma M_\nu{}\!^\alpha{}\!_\sigma -\tfrac{1}{4}g_{\mu\nu}M^{\alpha\sigma\beta}\,M_{\alpha\sigma\beta}  \right) \\
      &+ A^\alpha{}\!_{\mu\sigma} \, A_{\alpha\nu}{}\!^\sigma -\tfrac{1}{2} g_{\mu\nu}A^{\alpha\beta\sigma}\,A_{\alpha\beta\sigma} \quad .
  \endaligned
\end{equation}
If we use the total antisymmetry of $\bf A$, the antisymmetry of $\bf M$ in its first two indices, raise and lower some summed indices, and change some Greek indices to Latin indices, we obtain
\begin{equation}\label{E:w}
\aligned
T_{\mu\nu}=
         & \tfrac{1}{2}\left( J_{\mu i} h^i{}\!_\nu + J_{\nu i} h^i{}\!_\mu\right)
         -( M^\alpha{}\!_{\mu i} M_{\alpha\nu}{}\!^i - \tfrac{1}{4} g_{\mu\nu} M^{\alpha\sigma i} M_{\alpha\sigma i})\\
          &+A^{ij}{}\!_\mu A_{ij\nu} - \tfrac{1}{2} g_{\mu\nu} A^{ij\alpha} A_{ij\alpha}
\quad .
\endaligned
\end{equation}
The terms in Eq.~(\ref{E:w}) that involve $\bf A$ may be written in a more simple form.
We define a vector
$$
A^\mu = \tfrac{1}{3!}(-g)^{-1/2} e^{\mu\alpha\beta\sigma} A_{\alpha\beta\sigma} \quad ,
$$
where $e^{\mu\alpha\beta\sigma}$ is the usual Levi-Civita tensor density of weight $+1$. Thus,
$$
\aligned
A_\nu&= A^\mu \, g_{\mu\nu}= \tfrac{1}{3!}(-g)^{-1/2}e^{\mu\alpha\beta\sigma}g_{\mu\nu}A_{\alpha\beta\sigma}
    = \tfrac{1}{3!}(-g)^{-1/2}e^{\mu\alpha\beta\sigma}g_{\mu\nu}\, g_{\alpha\rho}\, g_{\beta\gamma}\, g_{\sigma\tau}A^{\rho\gamma\tau}\\
   &= \tfrac{1}{3!}(-g)^{-1/2}\, g \, e_{\nu\rho\gamma\tau}A^{\rho\gamma\tau}=-\tfrac{1}{3!}(-g)^{1/2}e_{\nu\rho\gamma\tau}A^{\rho\gamma\tau} \quad .
\endaligned
$$
We see from the above that we have the equivalent expressions
\begin{equation}\label{E:MMMM}
A_\nu = -\tfrac{1}{3!}(-g)^{1/2}e_{\nu\rho\gamma\tau}A^{\rho\gamma\tau}
\end{equation}
and
$$
A_\nu = \tfrac{1}{3!}(-g)^{-1/2}e^{\mu\alpha\beta\sigma}g_{\mu\nu}A_{\alpha\beta\sigma} \, .
$$
From the second expression  for  $A_\nu$, we have
\begin{equation}\label{E:NNNN}
A_\mu = \tfrac{1}{3!}(-g)^{-1/2}e^{\kappa\alpha\beta\sigma}g_{\kappa\mu}A_{\alpha\beta\sigma} \, .
\end{equation}
By multiplying Eqs.~(\ref{E:MMMM}) and (\ref{E:NNNN}), we get
$$
A_\mu \, A_\nu =  -\tfrac{1}{36}e^{\kappa\alpha\beta\sigma} \, e_{\nu\rho\gamma\tau}\, g_{\kappa\mu}A_{\alpha\beta\sigma} \, A^{\rho\gamma\tau} \quad .
$$
We now express the product of Levi-Civita symbols as a determinant of Kronecker deltas (see. e.g. Weber [40]), and find  that
$$
A_\mu A_\nu = \tfrac{1}{2} A^{ij}{}\!_\mu A_{ij\nu} - \tfrac{1}{6} g_{\mu\nu} A^{ij\alpha} A_{ij\alpha} \,
$$
and if we multiply by $g^{\mu\nu}$ we easily find that
\begin{equation}\label{E:GGGG}
A^{ij\alpha}\, A_{ij\alpha}= -6A^\alpha  A_\alpha \quad .
\end{equation}
We find from the last two equations that
$$
A^{ij}{}\!_\mu \, A_{ij\nu}=2A_\mu A_\nu  - 2g_{\mu\nu} A^\alpha  A_\alpha \quad .
$$
Now, by substituting the above expressions for $A^{ij\alpha}\, A_{ij\alpha}$ and $A^{ij}{}\!_\mu \, A_{ij\nu}$ into Eq.~(\ref{E:w}), we obtain the more simple expression, which is our Eq.~(\ref{E:EE}) in Sec.~5.3.
$$
\aligned
T_{\mu\nu}=
         & \tfrac{1}{2}\left( J_{\mu i} h^i{}\!_\nu + J_{\nu i} h^i{}\!_\mu\right)
         -( M^\alpha{}\!_{\mu i} M_{\alpha\nu}{}\!^i - \tfrac{1}{4} g_{\mu\nu} M^{\alpha\sigma i} M_{\alpha\sigma i})\\
          &+2A_\mu A_\nu + g_{\mu\nu} A^\alpha A_\alpha
\quad .
\endaligned
$$\newline

\paragraph{\bf{Equations (\ref{E:MM}) and (\ref{E:NN})}} We have\newline
$$
\aligned
\gamma^{\mu i\nu} \gamma_{\mu\nu i}&= (A^{\mu i\nu} +M^{\mu i\nu})(A_{\mu\nu i} +M_{\mu\nu i})\\
                        &= A^{\mu i\nu}A_{\mu\nu i}+A^{\mu i\nu}M_{\mu\nu i}+M^{\mu i\nu}A_{\mu\nu i}+M^{\mu i\nu}M_{\mu\nu i}
\endaligned
$$
and, from Eq.~(\ref{E:o}), we see that $A^{\mu i\nu}M_{\mu\nu i}=M^{\mu i\nu}A_{\mu\nu i}=0.$ Thus,
$$
\gamma^{\mu i\nu} \gamma_{\mu\nu i}=-A^{\mu\nu i}A_{\mu\nu i}+M^{i\mu\nu}M_{\nu\mu i} \quad .
$$
From Eq.~(\ref{E:GGGG}) we have $-A^{\mu\nu i}A_{\mu\nu i}= 6A^\alpha A_\alpha$, and from Eq.~(\ref{E:p}) we have
$M^{i\mu\nu}M_{\nu\mu i}=\tfrac{1}{2}M^{\mu\nu i}M_{\mu\nu i}$. Hence
$$
\gamma^{\mu i\nu} \gamma_{\mu\nu i} = \tfrac{1}{2} M^{\mu\nu i} M_{\mu\nu i} + 6 A^\alpha  A_\alpha  \,  ,
$$
as claimed above Eq.~(\ref{E:NN}).
\begin{center}
REFERENCES
\end{center}
1.\;Einstein, A.: Autobiographical notes. in Albert Einstein: Philosopher-\newline
\hphantom{B}Scientist, edited by P.~A. Schilpp, Harper \& Brothers, New York, Vol. I,\newline
\hphantom{B}p.~89 (1949)\newline
2.\;Dirac, P.~A.~M.: The Principles of Quantum Mechanics, Cambridge\newline
\hphantom{B}University Press, Preface to First Edition (1930) \newline
3.\;Pandres, D.,~Jr.: On forces and interactions between fields. J. Math. Phys.\newline
\hphantom{B}{\bf 3}, 602-607 (1962)\newline
4.\;Pandres, D.,~Jr.: Unification of gravitation and electromagnetism. Lett.\newline
\hphantom{B}Nuovo Cimento {\bf 8}, 595-599 (1973) \newline
5.\;Pandres, D.,~Jr.: Gravitation and electromagnetism. Found. Phys. {\bf 7},\newline
\hphantom{B}421-430 (1977) \newline
6.\;Feynman, R.~P., Hibbs, A.~R.: Quantum Mechanics and Path Integrals,
\hphantom{B}McGraw Hill (1965) \newline
7.\;Witten, E.: in Superstrings: A Theory of Everything, edited by P.~C.~W.\newline
\hphantom{B}Davis, J. Brown: Cambridge University Press, p.~90 (1988)\newline
8.\;Eddington, A.~S.: The Mathematical Theory of Relativity, 2nd Ed.,\newline
\hphantom{B}Cambridge University Press, p.~225 (1924)\newline
9.\;Penrose, R.: Int. J. Theor. Phys. {\bf 1}, 61 (1968)\newline
10.\;Penrose, R., MacCallum, A.~H.: Phys. Rep. {\bf 6C}, 242 (1972) and\newline
\hphantom{B}references contained therein\newline
11.\;Finkelstein, D.: Phys. Rev. {\bf 184}, 1261 (1969)\newline
12.\;Finkelstein, D.: Phys. Rev. D {\bf 5}, 321\newline
13.\;Finkelstein, D., Frye, G., Susskind, L.: Phys. Rev. D {\bf 9}, 2231 (1974)\newline
\hphantom{B}and references contained therein\newline
14.\;Bergmann, P.~G., Komar, A.: The coordinate group symmetries of gen-\newline
\hphantom{B}eral relativity, Int. J. Theor. Phys. {\bf 5}, 15-28 (1972)\newline
15.\;Gambini, R., Trias, A.: Geometrical origin of gauge theories, Phys. Rev.\newline
\hphantom{B}D15 {\bf 23}, 553-555 (1981)\newline
16.\;Connes, A. ``Noncommutative differential geometry and the structure\newline
\hphantom{B}of space time,'' Prepared for NATO Advance Study Institute on Quantum
\hphantom{B}Fields and Quantum Space Time, Cargese, France (1996)\newline
17.\;Crane, L. ``What Is the mathematical structure of quantum space-time?'',\newline
\hphantom{B}Los Alamos Preprints (LANL Number 0706.4452v1), June 29, 2007.\newline
18.\;Mandelstam, S.: Quantum electrodynamics without potentials, Ann.\newline
\hphantom{B}Phys., N.Y. {\bf 19}, 1-24 (1962)\newline
19.\;Pandres, D.,~Jr.: Quantum unified field theory from enlarged coordinate\newline
\hphantom{B}transformation group. Phys. Rev. D, {\bf 24}, 1499-1508 (1981)\newline
20.\;Pauli, W.: Theory of Relativity, Pergamon Press, Oxford, p.~5 (1958)
21.\;Pandres, D.,~Jr.: Quantum unified field theory from enlarged coordinate\newline
\hphantom{B}transformation group II. Phys. Rev. D, {\bf 30}, 317-324 (1984) \newline
22.\;Synge, J.~L.:  Relativity: The General Theory, North-Holland Publishing\newline
\hphantom{B}Company, Amsterdam, p.~117, 357 (1960)\newline
23.\;Eisenhart, L.~P.:  Riemannian Geometry, Princeton University Press\newline
\hphantom{B}(1964) \newline
24.\;Taub, A.~H.: in Perspectives in Geometry and Relativity, edited by\newline
\hphantom{B}Hoffmann, B., Indiana University Press, Bloomington, p.~360 (1966)\newline
25.\;Schr\"odinger, E.: Space-Time Structure, Cambridge University Press,\newline
\hphantom{B}p.~97,~99 (1960)\newline
26.\;Pandres, D.,~Jr.: Unified gravitational and Yang-Mills fields. Int. J.\
\hphantom{B}Theor. Phys. {\bf 34}, 733-759 (1995)\newline
27.\;Pandres, D.,~Jr.: Gravitational and electroweak unification. Int. J.\newline
\hphantom{B}Theor. Phys. {\bf 38}, 1783-1805 (1999)  \newline
28.\;Salam, A: in Proceedings of the 8th Nobel Symposium on Elementary\newline
\hphantom{B}Particle Theory, N. Svartholm, ed., Almquist Forlag, Stockholm, 367 (1967)\newline
29.\;Weinberg, S.: A model of leptons, Phys. Rev. Lett. {\bf 19}, 1264 (1967)\newline
30.\;Yang, C.N., Mills, R.L.: Conservation of isotopic spin and isotopic gauge\newline
\hphantom{B}invariance, Phys. Rev. {\bf 96}, 191 (1954)

\newpage

\noindent
31.\;Nakahara, M.: Geometry, Topology and Physics, Adam Hilger, New\newline
\hphantom{B}York, p.~344 (1990)\newline
32.\;Moriyasu, K.: An Elementary Primer for Gauge Theory, World Scientific,\newline
\hphantom{B}Singapore, p.~52 (1983)\newline
33.\;Pandres, D.,~Jr, Green, E.~L.: Unified field theory from enlarged trans-\newline
\hphantom{B}formation group: The inconsistent Hamiltonian. Int. J. Theor. Phys. {\bf 42},
\hphantom{B}1849-1873 (2003)\newline
34.\;Bergmann, P.~G.,Goldberg, I.: Dirac bracket transformations in phase\newline
\hphantom{B}space, Phys. Rev. {\bf 98}, 531-538 (1955)  \newline
35.\;Dirac, P.~A~M.: Lectures on Quantum Mechanics, Academic Press, New\newline
\hphantom{B}York (1964)\newline
36.\;Sundermeyer, K.: Constrained Dynamics, Springer-Verlag, Berlin (1982)\newline
37.\;Weinberg S.: The Quantum Theory of Fields, Vol. I, Cambridge Univer-\newline
\hphantom{B}sity Press, (1995)\newline
38.\;Pandres, D.,~Jr.: Quantum geometry from coordinate transformations\linebreak
\hphantom{B}relating quantum observers. Int. J. Theor. Phys. {\bf 23}, 839-842 (1984)\newline
39.\;Everett, H.~III: Relative state formulation of quantum mechanics, Revs.\newline
\hphantom{B}Mod. Phys. {\bf 29}, 454-462 (1957)\newline
40.\;Weber, J.: General Relativity and Gravitational Waves, Interscience Pub-\newline
\hphantom{B}lishers, New York, p.~147 (1961)\newline

\end{document}

%% file: TABLE.tex

\input nine

\input eight

\catcode `\!=11
\catcode `\@=11





\let\!tacr=\\ 


\newdimen\LineThicknessUnit
\newdimen\StrutUnit
\newskip \InterColumnSpaceUnit
\newdimen\ColumnWidthUnit
\newdimen\KernUnit

\let\!taLTU=\LineThicknessUnit 
\let\!taCWU=\ColumnWidthUnit   
\let\!taKU =\KernUnit          

\newtoks\NormalTLTU
\newtoks\NormalTSU
\newtoks\NormalTICSU
\newtoks\NormalTCWU
\newtoks\NormalTKU

\NormalTLTU={1in \divide \LineThicknessUnit by 300 }
\NormalTSU ={\normalbaselineskip
  \divide \StrutUnit by 11 }  
\NormalTICSU={.5em plus 1fil minus .25em}  
\NormalTCWU ={.5em}
\NormalTKU  ={.5em}

\def\NormalTableUnits{%
  \LineThicknessUnit   =\the\NormalTLTU
  \StrutUnit           =\the\NormalTSU
  \InterColumnSpaceUnit=\the\NormalTICSU
  \ColumnWidthUnit     =\the\NormalTCWU
  \KernUnit            =\the\NormalTKU}

\NormalTableUnits



\newcount\@multicnt
\newcount\LineThicknessFactor
\newcount\StrutHeightFactor
\newcount\StrutDepthFactor
\newcount\InterColumnSpaceFactor
\newcount\ColumnWidthFactor
\newcount\KernFactor
\newcount\VspaceFactor

\LineThicknessFactor    =2
\StrutHeightFactor      =8
\StrutDepthFactor       =3
\InterColumnSpaceFactor =3
\ColumnWidthFactor      =10
\KernFactor             =1
\VspaceFactor           =2


\newcount\TracingKeys 
\newcount\TracingFormats  


\def\BeginTableParBox#1{%
  \vtop\bgroup
    \hsize=#1
    \normalbaselines
    \let~=\!ttTie
    \let\-=\!ttDH
    \the\EveryTableParBox}

\def\EndTableParBox{%
    \MakeStrut{0pt}{\StrutDepthFactor\StrutUnit}
  \egroup} 

\newtoks\EveryTableParBox
\EveryTableParBox={%
  \parindent=0pt
  \raggedright
  \rightskip=0pt plus 4em 
  \relax}


\newtoks\EveryTable
\newtoks\!taTableSpread


\newskip\LeftTabskip
\newskip\RightTabskip


\newcount\!taCountA
\newcount\!taColumnNumber
\newcount\!taRecursionLevel 

\newdimen\!taDimenA  
\newdimen\!taDimenB  
\newdimen\!taDimenC  
\newdimen\!taMinimumColumnWidth

\newtoks\!taToksA

\newtoks\!taPreamble
\newtoks\!taDataColumnTemplate
\newtoks\!taRuleColumnTemplate
\newtoks\!taOldRuleColumnTemplate
\newtoks\!taLeftGlue
\newtoks\!taRightGlue

\newskip\!taLastRegularTabskip

\newif\if!taDigit
\newif\if!taBeginFormat
\newif\if!taOnceOnlyTabskip



\def\TaBlE{%
  T\kern-.27em\lower.5ex\hbox{A}\kern-.18em B\kern-.1em
    \lower.5ex\hbox{L}\kern-.075em E}



{\catcode`\|=13 \catcode`\"=13
  \gdef\ActivateBarAndQuote{%
    \ifnum \catcode`\|=13
    \else
      \catcode`\|=13
      \def|{%
        \ifmmode
          \vert
        \else
          \char`\|
        \fi}%
    \fi
    \ifnum \catcode`\"=13
    \else
      \catcode`\"=13
      \def"{\char`\"}%
    \fi}}

{\catcode `\|=12 \catcode `\"=12

}


\def\!thMessage#1{\immediate\write16{#1}\ignorespaces}

\let\!thx=\expandafter

\def\!thGobble#1{}

\def\\{\let\!thSpaceToken= }\\

\def\!thHeight{height}
\def\!thDepth{depth}
\def\!thWidth{width}

\def\!thToksEdef#1=#2{%
  \edef\!ttemp{#2}%
  #1\!thx{\!ttemp}%
  \ignorespaces}


\def\!thStoreErrorMsg#1#2{%
  \toks0 =\!thx{\csname #2\endcsname}%
  \edef#1{\the\toks0 }}

\def\!thReadErrorMsg#1{%
  \!thx\!thx\!thx\!thGobble\!thx\string #1}

\def\!thError#1#2{%
  \begingroup
    \newlinechar=`\^^J%
    \edef\!ttemp{#2}%
    \errhelp=\!thx{\!ttemp}%
    \!thMessage{%
      ^^J\!thReadErrorMsg\!thErrorMsgA
      ^^J\!thReadErrorMsg\!thErrorMsgB}%
    \errmessage{#1}%
  \endgroup}

\!thStoreErrorMsg\!thErrorMsgA{%
  TABLE error; see manual for explanation.}
\!thStoreErrorMsg\!thErrorMsgB{%
  Type \space H <return> \space for immediate help.}

\def\!thGetReplacement#1#2{%
   \begingroup
     \!thMessage{#1}
     \endlinechar=-1
     \global\read16 to#2%
   \endgroup}


\def\!thLoop#1\repeat{%
  \def\!thIterate{%
    #1%
    \!thx \!thIterate
    \fi}%
  \!thIterate
  \let\!thIterate\relax}


\def\Smash{%
  \relax
  \ifmmode
    \expandafter\mathpalette
    \expandafter\!thDoMathVCS
  \else
    \expandafter\!thDoVCS
  \fi}

\def\!thDoVCS#1{%
  \setbox\z@\hbox{#1}%
  \!thFinishVCS}

\def\!thDoMathVCS#1#2{%
  \setbox\z@\hbox{$\m@th#1{#2}$}%
  \!thFinishVCS}

\def\!thFinishVCS{%
  \vbox to\z@{\vss\box\z@\vss}}






\def\!thSetDimen{%
  \ifnum \!tgCode=1
    \ifx \!tgValue\empty
      \!taDimenA \StrutHeightFactor\StrutUnit
      \advance \!taDimenA \StrutDepthFactor\StrutUnit
      \divide \!taDimenA 2
    \else
      \!taDimenA \!tgValue\StrutUnit
    \fi
  \else
    \!taDimenA \!tgValue
  \fi
  \!taDimenA=\!thSign\!taDimenA\relax
  %
  \ifmmode
    \expandafter\mathpalette
    \expandafter\!thDoMathRaise
  \else
    \expandafter\!thDoSimpleRaise
  \fi}

\def\!thDoSimpleRaise#1{%
  \setbox\z@\hbox{\raise \!taDimenA\hbox{#1}}%
  \!thFinishRaise} 

\def\!thDoMathRaise#1#2{%
  \setbox\z@\hbox{\raise \!taDimenA\hbox{$\m@th#1{#2}$}}%
  \!thFinishRaise}

\def\!thFinishRaise{%
  \ht\z@\z@
  \dp\z@\z@
  \box\z@}


\def\!thKernBack{%
  \kern -
  \ifnum \!tgCode=1
    \ifx \!tgValue\empty
      \the\KernFactor
    \else
      \!tgValue    
    \fi
    \KernUnit
  \else
    \!tgValue      
  \fi
  \ignorespaces}%

\def\Vspace{%
  \noalign
  \bgroup
  \!tgGetValue\!thVspace}

\def\!thVspace{%
  \vskip
    \ifnum \!tgCode=1
      \ifx \!tgValue\empty
        \the\VspaceFactor
      \else
        \!tgValue    
      \fi
      \StrutUnit
    \else
      \!tgValue      
    \fi
  \egroup} 







\def\BeginFormat{%
  \catcode`\|=12 
  \catcode`\"=12 
  \!taPreamble={}%
  \!taColumnNumber=0
  \skip0 =\InterColumnSpaceUnit
  \multiply\skip0 \InterColumnSpaceFactor
  \divide\skip0 2
  \!taRuleColumnTemplate=\!thx{%
    \!thx\tabskip\the\skip0 }%
  \!taLastRegularTabskip=\skip0
  \!taOnceOnlyTabskipfalse
  \!taBeginFormattrue 
  \def\!tfRowOfWidths{}
  \ReadFormatKeys}

\def\!tfSetWidth{%
  \ifx \!tfRowOfWidths \empty  
    \ifnum \!taColumnNumber>0  
      \begingroup              
         \!taCountA=1          
         \aftergroup \edef \aftergroup \!tfRowOfWidths \aftergroup {%
           \aftergroup &\aftergroup \omit
           \!thLoop
             \ifnum \!taCountA<\!taColumnNumber
             \advance\!taCountA 1
             \aftergroup \!tfAOAO
           \repeat
           \aftergroup }%
      \endgroup
    \fi
  \fi
  \ifx [\!ttemp 
    \!thx\!tfSetWidthText
  \else
    \!thx\!tfSetWidthValue
  \fi}

\def\!tfAOAO{%
  &\omit&\omit}

\def\!tfSetWidthText [#1]{
  \def\!tfWidthText{#1}%
  \ReadFormatKeys}

\def\!tfSetWidthValue{%
  \!taMinimumColumnWidth =
    \ifnum \!tgCode=1
      \ifx\!tgValue\empty 
        \ColumnWidthFactor
      \else
        \!tgValue
      \fi
      \ColumnWidthUnit
    \else
      \!tgValue
    \fi
  \def\!tfWidthText{}
  \ReadFormatKeys}

\def\!tfSetTabskip{%
  \ifnum \!tgCode=1
    \skip0 =\InterColumnSpaceUnit
    \multiply\skip0
      \ifx \!tgValue\empty
        \InterColumnSpaceFactor         
      \else
       \!tgValue                        
      \fi
  \else
    \skip0 =\!tgValue                   
  \fi
  \divide\skip0 by 2
  \ifnum\!taColumnNumber=0
    \!thToksEdef\!taRuleColumnTemplate={%
      \the\!taRuleColumnTemplate
      \tabskip \the\skip0 }
  \else
    \!thToksEdef\!taDataColumnTemplate={%
      \the\!taDataColumnTemplate
      \tabskip \the\skip0 }
  \fi
  \if!taOnceOnlyTabskip
  \else
    \!taLastRegularTabskip=\skip0 
  \fi                             
  \ReadFormatKeys}

\def\!tfSetVrule{%
  \!thToksEdef\!taRuleColumnTemplate={%
    \noexpand\hfil
    \noexpand\vrule
    \noexpand\!thWidth
    \ifnum \!tgCode=1
      \ifx \!tgValue\empty
        \the\LineThicknessFactor      
      \else
        \!tgValue                     
      \fi
      \!taLTU                         
    \else
      \!tgValue                       
    \fi
    ####%
    \noexpand\hfil
    \the\!taRuleColumnTemplate}       
  \!tfAdjoinPriorColumn}

\def\!tfSetAlternateVrule{%
  \afterassignment\!tfSetAlternateA
  \toks0 =}                           

\def\!tfSetAlternateA{%
  \!thToksEdef\!taRuleColumnTemplate={%
    \the\toks0 \the\!taRuleColumnTemplate} 
  \!tfAdjoinPriorColumn}

\def\!tfAdjoinPriorColumn{%
  \ifnum \!taColumnNumber=0
    \!taPreamble=\!taRuleColumnTemplate 
    \ifnum \TracingFormats>0
      \!tfShowRuleTemplate
    \fi
  \else
    \ifx\!tfRowOfWidths\empty  
    \else
      \!tfUpdateRowOfWidths
    \fi
    \!thToksEdef\!taDataColumnTemplate={%
      \the \!taLeftGlue
      \the \!taDataColumnTemplate
      \the \!taRightGlue}
    \ifnum \TracingFormats>0
      \!tfShowTemplates
    \fi
    \!thToksEdef\!taPreamble={%
      \the\!taPreamble
      &
      \the\!taDataColumnTemplate
      &
      \the\!taRuleColumnTemplate}
  \fi
%
  \advance \!taColumnNumber 1
  \if!taOnceOnlyTabskip
    \!thToksEdef\!taDataColumnTemplate={%
       ####\tabskip \the\!taLastRegularTabskip}
  \else
    \!taDataColumnTemplate{##}%
  \fi
  \!taRuleColumnTemplate{}
  \!taLeftGlue{\hfil}
  \!taRightGlue{\hfil}%
  \!taMinimumColumnWidth=0pt
  \def\!tfWidthText{}%
  \!taOnceOnlyTabskipfalse    
  \ReadFormatKeys}

\def\!tfUpdateRowOfWidths{%
  \ifx \!tfWidthText\empty
  \else 
    \!tfComputeMinColWidth
  \fi
  \edef\!tfRowOfWidths{%
    \!tfRowOfWidths
    &%
    \omit                                  
    \ifdim \!taMinimumColumnWidth>0pt
      \hskip \the\!taMinimumColumnWidth
    \fi
    &
    \omit}}                                

\def\!tfComputeMinColWidth{%
  \setbox0 =\vbox{%
    \ialign{
       \span\the\!taDataColumnTemplate\cr
       \!tfWidthText\cr}}%
  \!taMinimumColumnWidth=\wd0 }

\def\!tfShowRuleTemplate{%
  \!thMessage{}
  \!thMessage{TABLE FORMAT}
  \!thMessage{Column: Template}
  \!thMessage{%
    \space *c: ##\tabskip \the\LeftTabskip}
  \!taOldRuleColumnTemplate=\!taRuleColumnTemplate}

\def\!tfShowTemplates{%
  \!thMessage{%
    \space \space r: \the\!taOldRuleColumnTemplate}
  \!taOldRuleColumnTemplate=\!taRuleColumnTemplate
  \!thMessage{%
    \ifnum \!taColumnNumber<10
      \space
    \fi
    \the\!taColumnNumber c: \the\!taDataColumnTemplate}
  \ifdim\!taMinimumColumnWidth>0pt
    \!thMessage{%
      \space \space w: \the\!taMinimumColumnWidth}
  \fi}

\def\!tfFinishFormat{%
  \ifnum \TracingFormats>0
    \!thMessage{%
      \space \space r: \the\!taOldRuleColumnTemplate
        \tabskip \the\RightTabskip}%
    \!thMessage{%
      \space *c: ##\tabskip 0pt}
  \fi
  \ifnum \!taColumnNumber<2
    \!thError{%
      \ifnum \!taColumnNumber=0
        No
      \else
        Only 1
      \fi
      "|"}%
      {\!thReadErrorMsg\!tfTooFewBarsA
       ^^J\!thReadErrorMsg\!tfTooFewBarsB
       ^^J\!thReadErrorMsg\!tkFixIt}%
  \fi
  \!thToksEdef\!taPreamble={%
    ####\tabskip\LeftTabskip
    &
    \the\!taPreamble \tabskip\RightTabskip
    &
    ####\tabskip 0pt \cr}
  \ifnum \TracingFormats>1
    \!thMessage{Preamble=\the\!taPreamble}
  \fi
  \ifnum \TracingFormats>2
    \!thMessage{Row Of Widths="\!tfRowOfWidths"}
  \fi
  \!taBeginFormatfalse 
  \catcode`\|=13
  \catcode`\"=13
  \!ttDoHalign}

\!thStoreErrorMsg\!tfTooFewBarsA{%
  There must be at least 2 "|"'s (and/or "\string \|"'s)}
\!thStoreErrorMsg\!tfTooFewBarsB{%
  between \string\BeginFormat\space and \string\EndFormat\space (or ".").}

\def\ReFormat[{%
  \omit
  \!taDataColumnTemplate{##}%
  \!taLeftGlue{}%
  \!taRightGlue{}%
  \catcode`\|=12  
  \catcode`\"=12  
  \ReadFormatKeys}

\def\!tfEndReFormat{%
  \ifnum \TracingFormats>0
    \!thMessage{ReF:
       \the\!taLeftGlue
       \hbox{\the\!taDataColumnTemplate}
       \the\!taRightGlue}
  \fi
  \catcode`\|=13
  \catcode`\"=13
  \!tfReFormat}

\def\!tfReFormat#1{%
  \the \!taLeftGlue
  \vbox{%
    \ialign{%
      \span\the\!taDataColumnTemplate\cr
       #1\cr}}%
  \the \!taRightGlue}







\def\!tgGetValue#1{%
  \def\!tgReturn{#1}
  \futurelet\!ttemp\!tgCheckForParen}

\def\!tgCheckForParen{%
  \ifx\!ttemp (%
    \!thx \!tgDoParen
  \else
    \!thx \!tgCheckForSpace
  \fi}

\def\!tgDoParen(#1){%
  \def\!tgCode{2}%
  \def\!tgValue{#1}
  \!tgReturn}

\def\!tgCheckForSpace{%
  \def\!tgCode{1}%
  \def\!tgValue{}
  \ifx\!ttemp\!thSpaceToken
    \!thx \!tgReturn        
  \else
    \!thx \!tgCheckForDigit
  \fi}

\def\!tgCheckForDigit{%
  \!taDigitfalse
  \ifx 0\!ttemp
    \!taDigittrue
  \else
    \ifx 1\!ttemp
      \!taDigittrue
    \else
      \ifx 2\!ttemp
        \!taDigittrue
      \else
        \ifx 3\!ttemp
          \!taDigittrue
        \else
          \ifx 4\!ttemp
            \!taDigittrue
          \else
            \ifx 5\!ttemp
              \!taDigittrue
            \else
              \ifx 6\!ttemp
                \!taDigittrue
              \else
                \ifx 7\!ttemp
                  \!taDigittrue
                \else
                  \ifx 8\!ttemp
                    \!taDigittrue
                  \else
                    \ifx 9\!ttemp
                      \!taDigittrue
                    \fi
                  \fi
                \fi
              \fi
            \fi
          \fi
        \fi
      \fi
    \fi
  \fi
  \if!taDigit
    \!thx \!tgGetNumber
  \else
    \!thx \!tgReturn
  \fi}

\def\!tgGetNumber{%
  \afterassignment\!tgGetNumberA
  \!taCountA=}
\def\!tgGetNumberA{%
  \edef\!tgValue{\the\!taCountA}%
  \!tgReturn}


\def\!tgSetUpParBox{%
  \edef\!ttemp{%
    \noexpand \ReadFormatKeys
    b{\noexpand \BeginTableParBox{%
      \ifnum \!tgCode=1
        \ifx \!tgValue\empty
          \the\ColumnWidthFactor
        \else
          \!tgValue    
        \fi
        \!taCWU        
      \else
        \!tgValue      
      \fi}}}%
  \!ttemp
  a{\EndTableParBox}}

\def\!tgInsertKern{%
  \edef\!ttemp{%
    \kern
    \ifnum \!tgCode=1
      \ifx \!tgValue\empty
        \the\KernFactor
      \else
        \!tgValue    
      \fi
      \!taKU         
    \else
      \!tgValue      
    \fi}%
  \edef\!ttemp{%
    \noexpand\ReadFormatKeys
    \ifh@            
      b{\!ttemp}
    \fi
    \ifv@            
      a{\!ttemp}
    \fi}%
  \!ttemp}




\def\NewFormatKey#1{%
  \!thx\def\!thx\!ttempa\!thx{\string #1}%
  \!thx\def\!thx\!ttempb\!thx{\csname !tk<\!ttempa>\endcsname}%
  \ifnum \TracingKeys>0
    \!tkReportNewKey
  \fi
  \!thx\ifx \!ttempb \relax
    \!thx\!tkDefineKey
  \else
    \!thx\!tkRejectKey
  \fi}

\def\!tkReportNewKey{%
  \!taToksA\!thx{\!ttempa}%
  \!thMessage{NEW KEY: "\the\!taToksA"}}

\def\!tkDefineKey{%
  \!thx\def\!ttempb}%

\def\!tkRejectKey{%
    \!taToksA\!thx{\!ttempa}%
    \!thError{Key letter "\the\!taToksA" already used}
      {\!thReadErrorMsg\!tkFixIt}
    \def\!tkGarbage}%

\!thStoreErrorMsg\!tkFixIt{%
  You'd better type \space 'E' \space and fix your file.}


\def\ReadFormatKeys#1{%
  \!thx\def\!thx\!ttempa\!thx{\string #1}%
  \!thx\def\!thx\!ttempb\!thx{\csname !tk<\!ttempa>\endcsname}%
  \ifnum \TracingKeys>1
    \!tkReportKey
  \fi
  \!thx\ifx \!ttempb\relax
    \!thx\!tkReplaceKey
  \else
    \!thx\!ttempb
  \fi}

\def\!tkReportKey{%
  \!taToksA\!thx{\!ttempa}%
  \!thMessage{KEY: "\the\!taToksA"}}

\def\!tkReplaceKey{%
  \!taToksA\!thx{\!ttempa}%
  \!thError {Undefined format key "\the\!taToksA"}
    {\!thReadErrorMsg\!tkUndefined ^^J\!thReadErrorMsg\!tkBadKey}
  \!tkReplaceKeyA}

\def\!tkReplaceKeyA{%
  \!thGetReplacement{\!thReadErrorMsg\!tkReplace}\!tkReplacement
  \!thx\ReadFormatKeys\!tkReplacement}

\!thStoreErrorMsg\!tkUndefined{%
  The format key in " "'s on the next to top line is undefined.}
\!thStoreErrorMsg\!tkBadKey{%
  Type \space E \space to quit now, or
  \space<CR> \space and respond to next prompt.}
\!thStoreErrorMsg\!tkReplace{%
  Type \space<replacement key><CR> \space,
   or simply \space<CR> \space to skip offending key:}


\NewFormatKey b#1{%
  \!thx\!tkJoin\!thx{\the\!taDataColumnTemplate}{#1}%
  \ReadFormatKeys}

\def\!tkJoin#1#2{%
  \!taDataColumnTemplate{#2#1}}%

\NewFormatKey a#1{%
  \!taDataColumnTemplate\!thx{\the\!taDataColumnTemplate #1}%
  \ReadFormatKeys}

\NewFormatKey \{{%
  \!taDataColumnTemplate=\!thx{\!thx{\the\!taDataColumnTemplate}}%
  \ReadFormatKeys}

\NewFormatKey *#1#2{%
  \!taCountA=#1\relax
  \!taToksA={}%
  \!thLoop
    \ifnum \!taCountA > 0
    \!taToksA\!thx{\the\!taToksA #2}%
    \advance\!taCountA -1
  \repeat
  \!thx\ReadFormatKeys\the\!taToksA}


\NewFormatKey \LeftGlue#1{%
  \!taLeftGlue{#1}%
  \ReadFormatKeys}

\NewFormatKey \RightGlue#1{%
  \!taRightGlue{#1}%
  \ReadFormatKeys}

\NewFormatKey c{%
  \ReadFormatKeys
  \LeftGlue\hfil
  \RightGlue\hfil}

\NewFormatKey l{%
  \ReadFormatKeys
  \LeftGlue{}   
  \RightGlue\hfil}

\NewFormatKey r{%
  \ReadFormatKeys
  \LeftGlue\hfil
  \RightGlue{}}

\NewFormatKey k{%
  \h@true
  \v@true
  \!tgGetValue{\!tgInsertKern}}

\NewFormatKey i{%
  \h@true
  \v@false
  \!tgGetValue{\!tgInsertKern}}

\NewFormatKey j{%
  \h@false
  \v@true
  \!tgGetValue{\!tgInsertKern}}


\NewFormatKey n{%
  \def\!tnStyle{}%
   \futurelet\!tnext\!tnTestForBracket}

\NewFormatKey N{%
  \def\!tnStyle{$}%
   \futurelet\!tnext\!tnTestForBracket}


\NewFormatKey m{%
  \ReadFormatKeys b$ a$}

\NewFormatKey M{%
  \ReadFormatKeys \{ b{$\displaystyle} a$}

\NewFormatKey \m{%
  \ReadFormatKeys l b{{}} m}

\NewFormatKey \M{%
  \ReadFormatKeys l b{{}} M}

\NewFormatKey f#1{%
  \ReadFormatKeys b{#1}}

\NewFormatKey B{%
  \ReadFormatKeys f\bf}

\NewFormatKey I{%
  \ReadFormatKeys f\it}

\NewFormatKey S{%
  \ReadFormatKeys f\sl}

\NewFormatKey R{%
  \ReadFormatKeys f\rm}

\NewFormatKey T{%
  \ReadFormatKeys f\tt}

\NewFormatKey p{%
  \!tgGetValue{\!tgSetUpParBox}}


\NewFormatKey w{%
  \!tkTestForBeginFormat w{\!tgGetValue{\!tfSetWidth}}}


\NewFormatKey s{%
  \!taOnceOnlyTabskipfalse    
  \!tkTestForBeginFormat t{\!tgGetValue{\!tfSetTabskip}}}

\NewFormatKey o{%
  \!taOnceOnlyTabskiptrue
  \!tkTestForBeginFormat o{\!tgGetValue{\!tfSetTabskip}}}


\NewFormatKey |{%
  \!tkTestForBeginFormat |{\!tgGetValue{\!tfSetVrule}}}

\NewFormatKey \|{%
  \!tkTestForBeginFormat \|{\!tfSetAlternateVrule}}


\NewFormatKey .{%
  \!tkTestForBeginFormat.{\!tfFinishFormat}}

\NewFormatKey \EndFormat{%
  \!tkTestForBeginFormat\EndFormat{\!tfFinishFormat}}

\NewFormatKey ]{%
  \!tkTestForReFormat ] \!tfEndReFormat}


\def\!tkTestForBeginFormat#1#2{%
  \if!taBeginFormat
    \def\!ttemp{#2}%
    \!thx \!ttemp
  \else
    \toks0={#1}%
    \toks2=\!thx{\string\ReFormat}%
    \!thx \!tkImproperUse
  \fi}

\def\!tkTestForReFormat#1#2{%
  \if!taBeginFormat
    \toks0={#1}%
    \toks2=\!thx{\string\BeginFormat}%
    \!thx \!tkImproperUse
  \else
    \def\!ttemp{#2}%
    \!thx \!ttemp
  \fi}

\def\!tkImproperUse{%
  \!thError{\!thReadErrorMsg\!tkBadUseA "\the\toks0 "}%
    {\!thReadErrorMsg\!tkBadUseB \the\toks2 \space command.
    ^^J\!thReadErrorMsg\!tkBadKey}%
  \!tkReplaceKeyA}

\!thStoreErrorMsg\!tkBadUseA{Improper use of key }
\!thStoreErrorMsg\!tkBadUseB{%
  The key mentioned above can't be used in a }




\def\!tnTestForBracket{%
  \ifx [\!tnext
    \!thx\!tnGetArgument
  \else
    \!thx\!tnGetCode
  \fi}

\def\!tnGetCode#1 {
  \!tnConvertCode #1..!}

\def\!tnConvertCode #1.#2.#3!{%
  \begingroup
    \aftergroup\edef \aftergroup\!ttemp \aftergroup{%
      \aftergroup[%
      \!taCountA #1
      \!thLoop
        \ifnum \!taCountA>0
        \advance\!taCountA -1
        \aftergroup0
      \repeat
      \def\!ttemp{#3}%
      \ifx\!ttemp \empty
      \else
        \aftergroup.
        \!taCountA #2
        \!thLoop
          \ifnum \!taCountA>0
          \advance\!taCountA -1
          \aftergroup0
        \repeat
      \fi
      \aftergroup]\aftergroup}%
    \endgroup\relax
    \!thx\!tnGetArgument\!ttemp}

\def\!tnGetArgument[#1]{%
  \!tnMakeNumericTemplate\!tnStyle#1..!}

\def\!tnMakeNumericTemplate#1#2.#3.#4!{
  \def\!ttemp{#4}%
  \ifx\!ttemp\empty
    \!taDimenC=0pt
  \else
    \setbox0=\hbox{\m@th #1.#3#1}%
    \!taDimenC=\wd0
  \fi
  \setbox0 =\hbox{\m@th #1#2#1}%
  \!thToksEdef\!taDataColumnTemplate={%
    \noexpand\!tnSetNumericItem
    {\the\wd0 }%
    {\the\!taDimenC}%
    {#1}%
    \the\!taDataColumnTemplate}  
  \ReadFormatKeys}

\def\!tnSetNumericItem #1#2#3#4 {
  \!tnSetNumericItemA {#1}{#2}{#3}#4..!}

\def\!tnSetNumericItemA #1#2#3#4.#5.#6!{%
  \def\!ttemp{#6}%
  \hbox to #1{\hss \m@th #3#4#3}%
  \hbox to #2{%
    \ifx\!ttemp\empty
    \else
       \m@th #3.#5#3%
    \fi
    \hss}}




\def\MakeStrut#1#2{%
  \vrule width0pt height #1 depth #2}

\def\StandardTableStrut{%
  \MakeStrut{\StrutHeightFactor\StrutUnit}
    {\StrutDepthFactor\StrutUnit}}

\def\AugmentedTableStrut#1#2{%
  \dimen@=\StrutHeightFactor\StrutUnit
  \advance\dimen@ #1\StrutUnit
  \dimen@ii=\StrutDepthFactor\StrutUnit
  \advance\dimen@ii #2\StrutUnit
  \MakeStrut{\dimen@}{\dimen@ii}}

\def\Enlarge#1#2{
  \!taDimenA=#1\relax
  \!taDimenB=#2\relax
  \let\!TsSpaceFactor=\empty
  \ifmmode
    \!thx \mathpalette
    \!thx \!TsEnlargeMath
  \else
    \!thx \!TsEnlargeOther
  \fi}

\def\!TsEnlargeOther#1{%
  \ifhmode
    \setbox\z@=\hbox{#1%
      \xdef\!TsSpaceFactor{\spacefactor=\the\spacefactor}}%
  \else
    \setbox\z@=\hbox{#1}%
  \fi
  \!TsFinishEnlarge}

\def\!TsEnlargeMath#1#2{%
  \setbox\z@=\hbox{$\m@th#1{#2}$}%
  \!TsFinishEnlarge}

\def\!TsFinishEnlarge{%
  \dimen@=\ht\z@
  \advance \dimen@ \!taDimenA
  \ht\z@=\dimen@
  \dimen@=\dp\z@
  \advance \dimen@ \!taDimenB
  \dp\z@=\dimen@
  \box\z@ \!TsSpaceFactor{}}


\def\OpenUp#1#2{%
  \advance \StrutHeightFactor #1\relax
  \advance \StrutDepthFactor #2\relax}




\def\BeginTable{%
  \futurelet\!tnext\!ttBeginTable}

\def\!ttBeginTable{%
  \ifx [\!tnext
    \def\!tnext{\!ttBeginTableA}%
  \else
    \def\!tnext{\!ttBeginTableA[c]}%
  \fi
  \!tnext}

\def\!ttBeginTableA[#1]{%
  \if #1u
    \ifmmode
      \def\!ttEndTable{
        \relax}
    \else                   
      \bgroup
      \def\!ttEndTable{%
        \egroup}%
    \fi
  \else
    \hbox\bgroup $
    \def\!ttEndTable{%
      \egroup 
      $
      \egroup}
    \if #1t%
      \vtop
    \else
      \if #1b%
        \vbox
      \else
        \vcenter 
      \fi
    \fi
    \bgroup      
  \fi
  \advance\!taRecursionLevel 1 
  \let\!ttRightGlue=\relax  
  \everycr={}
  \ifnum \!taRecursionLevel=1
    \!ttInitializeTable
  \fi}

\bgroup
  \catcode`\|=13
  \catcode`\"=13
  \catcode`\~=13
  \gdef\!ttInitializeTable{%
    \let\!ttTie=~ 
    \let\!ttDH=\- 
    \catcode`\|=\active
    \catcode`\"=\active
    \catcode`\~=\active
    \def |{\unskip\!ttRightGlue&&}
    \def\|{\unskip\!ttRightGlue&\omit\!ttAlternateVrule}%
    \def"{\unskip\!ttRightGlue&\omit&}
    \def~{\kern .5em}
    \def\\{\!ttEndOfRow}%
    \def\-{\!ttShortHrule}%
    \def\={\!ttLongHrule}%
    \def\_{\!ttFullHrule}%
    \def\Left##1{##1\hfill\null}
    \def\Center##1{\hfill ##1\hfill\null}
    \def\Right##1{\hfill##1}%
    \the\EveryTable}
\egroup

\let\!ttRightGlue=\relax  

\def\!ttDoHalign{%
  \baselineskip=0pt \lineskiplimit=0pt \lineskip=0pt %
  \tabskip=0pt
  \halign \the\!taTableSpread \bgroup
   \span\the\!taPreamble
   \ifx \!tfRowOfWidths \empty
   \else
     \!tfRowOfWidths \cr %
   \fi}

\def\EndTable{%
  \egroup 
  \!ttEndTable}


\def\!ttEndOfRow{%
  \futurelet\!tnext\!ttTestForBlank}

\def\!ttTestForBlank{%
  \ifx \!tnext\!thSpaceToken  
    \!thx\!ttDoStandard
  \else
    \!thx\!ttTestForZero
  \fi}

\def\!ttTestForZero{%
  \ifx 0\!tnext
    \!thx \!ttDoZero
  \else
    \!thx \!ttTestForPlus
  \fi}

\def\!ttTestForPlus{%
  \ifx +\!tnext
    \!thx \!ttDoPlus
  \else
    \!thx \!ttDoStandard
  \fi}

\def\!ttDoZero#1{
  \cr}

\def\!ttDoPlus#1#2#3{
  \AugmentedTableStrut{#2}{#3}%
  \cr}

\def\!ttDoStandard{%
  \StandardTableStrut
  \cr}






\def\!ttAlternateVrule{%
  \!tgGetValue{\!ttAVTestForCode}}  

\def\!ttAVTestForCode{%
  \ifnum \!tgCode=2              
    \!thx\!ttInsertVrule         
  \else
    \!thx\!ttAVTestForEmpty
  \fi}

\def\!ttAVTestForEmpty{%
  \ifx \!tgValue\empty           
    \!thx\!ttAVTestForBlank
  \else
    \!thx\!ttInsertVrule         
  \fi}

\def\!ttAVTestForBlank{%
  \ifx \!ttemp\!thSpaceToken     
    \!thx\!ttInsertVrule
  \else
    \!thx\!ttAVTestForStar
  \fi}

\def\!ttAVTestForStar{%
  \ifx *\!ttemp                  
    \!thx\!ttInsertDefaultPR     
  \else
    \!thx\!ttGetPseudoVrule       
  \fi}

\def\!ttInsertVrule{%
  \hfil
  \vrule \!thWidth
    \ifnum \!tgCode=1
      \ifx \!tgValue\empty
        \LineThicknessFactor
      \else
        \!tgValue
      \fi
      \LineThicknessUnit
    \else
      \!tgValue
    \fi
  \hfil
  &}

\def\!ttInsertDefaultPR*{%
  \PseudoVrule    
  &}

\def\!ttGetPseudoVrule#1{%
  \toks0={#1}%
  #1&}

\def\PseudoVrule{}

%
%
\def\ifundefined#1{\expandafter\ifx\csname#1\endcsname\relax}
\def\!ttuse#1{%
  \ifnum #1>\@ne
    \omit
    \ifundefined{mscount}
       \@multicnt=#1
       \advance\@multicnt by \m@ne
       \advance\@multicnt by \@multicnt
       \!thLoop
         \ifnum\@multicnt>\@ne
         \sp@n %
       \repeat
    \else
       \mscount=#1
       \advance\mscount by \m@ne
       \advance\mscount by \mscount
       \!thLoop
         \ifnum\mscount>\@ne
         \sp@n %
       \repeat
    \fi
    \span
  \fi}

\def\!ttUse#1[{%
  \!ttuse{#1}%
  \ReFormat[}


\def\!ttFullHrule{%
  \noalign
  \bgroup
  \!tgGetValue{\!ttFullHruleA}}

\def\!ttFullHruleA{%
  \!ttGetHalfRuleThickness 
  \hrule \!thHeight \dimen0 \!thDepth \dimen0
  \penalty0 
  \egroup} 

\def\!ttShortHrule{%
  \omit
  \!tgGetValue{\!ttShortHruleA}}

\def\!ttShortHruleA{%
  \!ttGetHalfRuleThickness 
  \leaders \hrule \!thHeight \dimen0 \!thDepth \dimen0 \hfill
  \null    
  \ignorespaces}

\def\!ttLongHrule{%
  \omit\span\omit\span \!ttShortHrule}

\def\!ttGetHalfRuleThickness{%
  \dimen0 =
    \ifnum \!tgCode=1
      \ifx \!tgValue\empty
        \LineThicknessFactor
      \else
        \!tgValue    
      \fi
      \LineThicknessUnit
    \else
      \!tgValue      
    \fi
  \divide\dimen0 2 }



\def\WidenTableBy#1{%
  \ifdim #1=0pt
    \!taTableSpread={}%
  \else
    \!taTableSpread={spread #1}%
  \fi}

%


\def\JustLeft{%
  \omit \let\!ttRightGlue=\hfill}
\def\JustCenter{%
  \omit \hfill\null \let\!ttRightGlue=\hfill}

\let\\=\!tacr
\catcode`\!=12
\catcode`\@=12

%% file: nine.tex
\font\ninerm=cmr9
\font\sevenrm=cmr7
\font\sixrm=cmr6
\font\fiverm=cmr5
\font\ninei=cmmi9
\font\sixi=cmmi6
\font\fivei=cmmi6
\font\ninesy=cmsy9
\font\sixsy=cmsy6
\font\fivesy=cmsy5
\font\tenex=cmex10
\font\nineit=cmti9
\font\ninesl=cmsl9
\font\ninett=cmtt9
\font\ninebf=cmbx9
\font\sixbf=cmbx6
\font\fivebf=cmbx5

\def\ninepoint{\def\rm{\fam0\ninerm}
  \textfont0=\ninerm \scriptfont0=\sixrm \scriptscriptfont0=\fiverm
  \textfont1=\ninei \scriptfont1=\sixi \scriptscriptfont0=\fivei
  \textfont2=\ninesy \scriptfont2=\sixsy \scriptscriptfont2=\fivesy
  \textfont3=\tenex \scriptfont3=\tenex \scriptscriptfont3=\tenex
  \textfont\itfam=\nineit  \def\it{\fam\itfam\nineit}%
  \textfont\slfam=\ninesl  \def\sl{\fam\slfam\ninesl}%
  \textfont\ttfam=\ninett  \def\tt{\fam\ttfam\ninett}%
  \textfont\bffam=\ninebf  \scriptfont\bffam=\sixbf
   \scriptscriptfont\bffam=\fivebf  \def\bf{\fam\bffam\ninebf}%
  \normalbaselineskip=11pt
  \setbox\strutbox=\hbox{\vrule height8pt depth3pt width0pt}%
  \let\sc=\sevenrm  \normalbaselines\rm}

\def\ifundefined#1{\expandafter\ifx\csname#1\endcsname\relax}
\ifundefined{mscount}
  \renewcommand{\ninepoint}{
    \fontsize{9pt}{11}
    \selectfont
  }
\fi

%% file: eight.tex
\font\eightrm=cmr8
\font\sevenrm=cmr7
\font\sixrm=cmr6
\font\fiverm=cmr5
\font\eighti=cmmi8
\font\sixi=cmmi6
\font\fivei=cmmi6
\font\eightsy=cmsy8
\font\sixsy=cmsy6
\font\fivesy=cmsy5
\font\tenex=cmex10
\font\eightit=cmti8
\font\eightsl=cmsl8
\font\eighttt=cmtt8
\font\eightbf=cmbx8
\font\sixbf=cmbx6
\font\fivebf=cmbx5

\def\eightpoint{\def\rm{\fam0\eightrm}
  \textfont0=\eightrm \scriptfont0=\sixrm \scriptscriptfont0=\fiverm
  \textfont1=\eighti \scriptfont1=\sixi \scriptscriptfont0=\fivei
  \textfont2=\eightsy \scriptfont2=\sixsy \scriptscriptfont2=\fivesy
  \textfont3=\tenex \scriptfont3=\tenex \scriptscriptfont3=\tenex
  \textfont\itfam=\eightit  \def\it{\fam\itfam\eightit}%
  \textfont\slfam=\eightsl  \def\sl{\fam\slfam\eightsl}%
  \textfont\ttfam=\eighttt  \def\tt{\fam\ttfam\eighttt}%
  \textfont\bffam=\eightbf  \scriptfont\bffam=\sixbf
   \scriptscriptfont\bffam=\fivebf  \def\bf{\fam\bffam\eightbf}%
  \normalbaselineskip=9pt
  \setbox\strutbox=\hbox{\vrule height7pt depth2pt width0pt}%
  \let\sc=\sixrm  \normalbaselines\rm}

\def\ifundefined#1{\expandafter\ifx\csname#1\endcsname\relax}
\ifundefined{mscount}
  \renewcommand{\eightpoint}{
    \fontsize{8pt}{9.5}
    \selectfont
  }
\fi